\documentclass[aps,pra,amssymb,amsmath,eqsecnum,nofootinbib]{revtex4}
\usepackage{graphicx}

\newtheorem{definition}{Definition}[section]
\newtheorem{theorem}{Theorem}[section]
\newtheorem{proposition}{Proposition}[section]

\newtheorem{remark}{Remark}[section]
\newtheorem{axiom}{AXIOM}[section]

\newenvironment{hypothesis}{HP: \begin{center}} {\end{center}}
\newenvironment{thesis}{TH: \begin{center}} {\end{center}}
\newenvironment{proof}{\begin{center}PROOF: \end{center}} {$ \blacksquare $}
\newtheorem{example}{Example}[section]
\begin{document}
\title{The mathematical role of (commutative and noncommutative) infinitesimal random walks over (commutative and noncommutative) riemannian manifolds in Quantum Physics}
\author{Gavriel Segre}
\homepage{www.gavrielsegre.com}
\begin{abstract}
 Anderson's nonstandard construction of brownian motion as an
 infinitesimal random walk on the euclidean line is generalized to an Hausdorff
 riemannian manifold.

 A nonstandard Feynman-Kac formula holding on such an Hausdorff
 riemannian manifold is derived.

 Indications are given on how these (radically elementary) results could allow to formulate a
 nonstandard version of Stochastic Mechanics (avoiding both the
 explicitly discussed bugs of Internal Set Theory as well as the
 controversial renormalization of the stochastic
 action).

It is anyway remarked how this would contribute to hide the basic
feature of Quantum Mechanics, i.e. the noncommutativity of the
observables' algebra, whose structure is naturally captured  in
the language of Noncommutative Probability and Noncommutative
Geometry.

With this respect some preliminary consideration concerning the
notion of infinitesimal quantum random walk on a noncommutative
riemannian manifold, the notion obtained by the Sinha-Goswami's
definition of quantum brownian motion on a noncommutative
riemannian manifold replacing a continuous time interval with an
hyperfinite time interval, is presented.
\end{abstract}
\maketitle
\newpage
\tableofcontents
\newpage
\section{Feynman's physical intuition about path integrals} \label{sec:Feynman's physical intuition about path integrals}
Let us briefly recall Feyman's way of arriving to his
mathematically non rigorous but physically extraordinary
path-integral formulation of Quantum  Mechanics (see for instance
\cite{Feynman-48}, \cite{Feynman-Hibbs-65}, \cite{Schulman-81},
\cite{Kleinert-95}, \cite{Grosche-Steiner-98}):
\begin{equation} \label{eq:Feynman's equation}
    < q''  | \hat{U}_{t} |   q'  > \; = \; \int_{\left\{%
\begin{array}{ll}
    q( 0 ) = q',  \\
     q( t ) = q'' \\
\end{array}%
\right.      } [ d q(s) ] \exp ( i S[ q(s)]) \; \; \forall q', q''
\in \mathbb{R} , \forall t \in (0, + \infty )
\end{equation}
(where we  adopt, from here and beyond, a unit system in which $
\hbar = 1 $ and where $ S[q] := \int_{0}^{t} ds ( \frac{1}{2}
\dot{q}^{2} - V[q(s)]$ is the non-relativistic classical action
functional for a particle of unary mass living on the real line
under the influence of the conservative field force of energy
potential V) recasting it in the more usual euclidean expression:
\begin{equation} \label{eq:Feynman's euclidean equation}
    < q''  | \hat{U}_{t} |   q'  > \; = \; \int_{\left\{%
\begin{array}{ll}
    q( 0 ) = q',  \\
     q( t ) = q'' \\
\end{array}%
\right.      } [ d q(s) ] \exp ( - S_{E}[ q(s)]) \; \; \forall q',
q'' \in \mathbb{R} , \forall t \in (0, + \infty )
\end{equation}
obtained from eq.\ref{eq:Feynman's equation} through Wick's
rotation, i.e. prolonging analytically to complex values the time
parameter t and computing it on the positive imaginary axis.

The basic steps are:
\begin{enumerate}
    \item the introduction of $ n \in \mathbb{N}_{+} $  times:
\begin{equation}
    t_{k} \; := \;  k \epsilon \; \; k \in  \{ j \in \mathbb{N} \, : \,  j \leq   n \}
\end{equation}
where:
\begin{equation}
    \epsilon \; := \; \frac{ t }{n}
\end{equation}
    \item the adoption of the semi-group property of the euclidean time-evolution
    operator $ \hat{U} $ in order to write:
\begin{equation}
  < q''  | \hat{U}_{t} |   q'  > \; = \; \prod_{k=1}^{n-1} \int_{- \infty}^{+
   \infty} d q_{k} \prod_{k=1}^{n}  < q_{k}  | \hat{U}_{\epsilon} |   q_{k-1}  >
\end{equation}
where $ q_{0} := q' $ and $ q_{n} := q'' $.
    \item the observation that:
\begin{equation}
      < q_{k}  | \hat{U}_{\epsilon} |   q_{k-1}  > \; = \;
      ( \frac{1}{2 \pi \epsilon } )^{\frac{1}{2}} \exp ( -  \frac{ ( q_{k} - q_{k-1} )^{2} }{ 2 \epsilon
      } - \epsilon V(q_{k}) + O( \epsilon^{2} ) )
\end{equation}
    \item a mathematically meaningless
    limit for $ n \rightarrow + \infty $ (and hence $ \epsilon
    \rightarrow 0 )$ in which the mathematically meaningless
    object $ \lim_{n \rightarrow + \infty}  ( \frac{1}{2 \pi \epsilon }
    )^{\frac{n}{2}} \prod_{k=1}^{n-1} d q_{k} $ is replaced with
    the mathematically meaningless functional measure $ [ d q(t)
    ] $.
\end{enumerate}

\smallskip

Such a derivation is as more impressive for the complete lack of
mathematical meaning of it last step as well as it is impressive
for the geniality of the physical intuition underlying it, a
geniality resulting in the physically extraordinary path-integral
formulation of Quantum Mechanics, with no doubt one of the
greatest achievements of $20^{th} $-century's Theoretical Physics.

\newpage
\section{The conditional Wiener measure and the Feynman-Kac formula on the euclidean line}
Let us now briefly recall how eq. \ref{eq:Feynman's euclidean
equation} can be recasted in a mathematically meaningful form in
terms of the Wiener measure (see for instance
\cite{Glimm-Jaffe-87}, \cite{Simon-05}).

Denoted with $ dq $ the Lebesgue measure over the measurable space
$ ( \mathbb{R} , \mathcal{B} ( \mathbb{R} )) $ (where of course $
\mathcal{B} ( \mathbb{R} )$ is the Borel-$\sigma$-algebra
associated to the natural topology over $ \mathbb{R} $, namely the
topology induced by the metric  $ d_{euclidean}( q' , q'' ) := |
q' - q'' | $ ) let us introduce the Hilbert space $ \mathcal{H} :=
L^{2}( \mathbb{R} , dq ) $ and the strongly-continuous contraction
semigroup (see \cite{Reed-Simon-80}, \cite{Reed-Simon-75}) $ \{
T^{(0)}_{t} := \exp ( \frac{t}{2} \frac{d^{2}}{d q^{2}} )\}_{t \in
( 0 , + \infty)} $  of operators on $ \mathcal{H} $.

It may be easily proved that the integral kernel $ K^{(0)}_{t}(
    q', q'' ) $ of $  \{ T^{(0)}_{t} \}_{t \in ( 0 , +
\infty)} $:
\begin{equation}
    ( T^{(0)}_{t} f ) (q') \; =: \int_{-\infty}^{+\infty} d q'' \,  K^{(0)}_{t}(
    q', q'' ) f(q'') \; \; \forall q' \in
    \mathbb{R} , \forall t \in ( 0 , + \infty )
\end{equation}
is given by:
\begin{equation}
  K^{(0)}_{t}( q', q'' ) \; = \; \frac{1}{ ( 2 \pi
  t)^{\frac{1}{2}}} \exp ( - \frac{ ( q'' - q' )^{2} }{2 t} )
  \; \; \forall q', q'' \in
    \mathbb{R} , \forall t \in ( 0 , + \infty )
\end{equation}
Let us now observe that $   K^{(0)}_{t}( q', q'' ) $ satisfies the
following basic properties:
\begin{enumerate}
    \item
\begin{equation}
   K^{(0)}_{t}( q', q'' ) > 0  \; \; \forall q', q'' \in
    \mathbb{R} , \forall t \in ( 0 , + \infty )
\end{equation}
    \item
\begin{equation}
  \int_{- \infty}^{+ \infty}  d
   q'' \,  K^{(0)}_{t}( q', q'' ) \; = \; 1  \; \; \forall q' \in
    \mathbb{R} , \forall t \in ( 0 , + \infty )
\end{equation}
    \item
\begin{equation}
   K^{(0)}_{t_{1}+t_{2}}( q', q''' ) \; = \; \int_{- \infty}^{+ \infty} d
   q'' \,  K^{(0)}_{t_{1}}( q', q'' )  K^{(0)}_{t_{2}}( q'', q''' ) \; \; \forall q', q''' \in
    \mathbb{R} , \forall t_{1} , t_{2} \in ( 0 , + \infty )
\end{equation}
\end{enumerate}
that guarantee that  $   K^{(0)}_{t}( q', q'' ) $ is the
transition probability kernel of a Markovian stochastic process $
w_{t} $ that one defines as the brownian motion  over the metric
space $ ( \mathbb{R} , d_{euclidean} ) $.

Introduced furthermore the functional space:
\begin{equation}
  \mathcal{C} ( q', 0 ; q'' , t ) \; := \{ q : [ 0 , t ]
  \mapsto \mathbb{R} \; d_{euclidean}-continuous \; : q(0) = q' ,  q(t) =
  q'' \} \; \; q' , q'' \in \mathbb{R} , t \in ( 0 , +
  \infty )
\end{equation}
\footnote{where $  d_{euclidean}$-continuity, i.e continuity
w.r.t. to the Borel $ \sigma $-algebra associated to the metric
topology $ \mathcal{B} ( \mathbb{R} ) $ induced by the metric $
d_{euclidean} $, is the usual notion of continuity of Real
Analysis.} and the cylinder sets of $  \mathcal{C} ( q', 0 ; q'' ,
t ) $:
\begin{multline}
    \Gamma ( q' , 0 ; \{ B_{k} ; t_{k} \}_{k=1}^{n-1} ;
    q'' , t ) \; := \; \{ q \in \mathcal{C} ( q', 0 ; q'' ,
t ) : q(t_{k}) \in B_{k} \\
     \forall k \in \{ j \in \mathbb{N} \, : 1 \leq j \leq  n -1
    \} \} \; \; 0 < \cdots < t_{k} < t_{k+1}   < \cdots <  t ,  B_{k} \in
    \mathcal{B}( \mathbb{R} )  \, \forall k \in \{ j \in \mathbb{N} \, : 1 \leq j \leq
    n-1
    \} \, , \,  n \in \mathbb{N}_{+}
\end{multline}
one defines the value of the conditional Wiener measure on
cylinder sets as:
\begin{equation}
    W [  \Gamma ( q' , 0 ; \{ B_{k} ; t_{k} \}_{k=1}^{n-1} ;
    q'' , t ) ] \; := \; \prod_{k=1}^{n-1} \int_{ B_{k}} d q_{k} \prod_{k=1}^{n}
    K^{(0)}_{t_{k}-t_{k-1}} ( q_{k-1} , q_{k} )
\end{equation}
where $ q_{0} := q' $ and $ q_{n} := q''$.

Then it may be proved that:
\begin{enumerate}
    \item the conditional Wiener measure $ W_{t} ( q' , q'' )$ is countably-additive on
    the cylinder sets of $  \mathcal{C} ( q', 0 ; q'' , t ) $
    \item by Kolmogorov Reconstruction Theorem it has a unique
    extension to the Borel subsets of $  \mathcal{C} ( q', 0 ; q'' , t )
    $ that one defines to be the  conditional Wiener measure $ W_{t} ( q' , q'' )$
\end{enumerate}

Given a potential energy V bounded from below let us introduce the
one-parameter family of operators $ \{ T^{(V)}_{t} :=  \exp [ t (
\frac{1}{2} \frac{d^{2}}{d q^{2}} - V(q) )] \}_{t \in ( 0 , +
\infty )} $ and the associated integral kernel:
\begin{equation}
    ( T^{(V)}_{t} f ) (q') \; =: \int_{-\infty}^{+\infty} d q'' \,  K^{(V)}_{t}(
    q', q'' ) f(q'') \; \; \forall q' \in
    \mathbb{R} , \forall t \in ( 0 , + \infty )
\end{equation}

The application of the Trotter-Kato formula:
\begin{equation}
  \exp (  \frac{1}{2} \frac{d^{2}}{d q^{2}} - V(q) ) \; = \;
  s-\lim_{n \rightarrow + \infty} [ \exp (  \frac{1}{2 n}  \frac{d^{2}}{d
  q^{2}}) \exp ( - \frac{V}{n} )]^{n}
\end{equation}
allows to derive the following:
\begin{theorem} \label{th:Feynman-Kac formula on the euclidean line}
\end{theorem}
\emph{Feynman-Kac formula on the euclidean line:}
\begin{equation}
   K^{(V)}_{t}( q', q'' ) \; = \; \int_{ \mathcal{C} ( q', 0 ; q'' , t
   )} d W_{t} ( q' , q'' ) \exp ( - \int_{0}^{t} d s V[q(s)]
   )
   \; \;  \forall q' , q'' \in \mathbb{R} \, , \,  \forall  t \in ( 0 , +
  \infty )
\end{equation}

\begin{remark} \label{rem:on the underlying geometry}
\end{remark}
Though often called simply the brownian motion on $ \mathbb{R} $
the Wiener process has to be more precisely called the brownian
motion on the one-dimensional euclidean manifold $ ( \mathbb{R} ,
\delta_{euclidean} ) $ where $ \delta_{euclidean} $ is the
euclidean riemannian metric over $ \mathbb{R} $.

In fact the definition of the brownian motion involves the choice
of a particular $ \sigma$-algebra over $ \mathbb{R} $ that is the
Borel-$\sigma$-algebra associated to the topology over $
\mathbb{R} $ induced by the metric $ d_{euclidean} $ induced by
the norm on $ \mathbb{R} $ induced by the euclidean
riemannian-metric $ \delta_{euclidean} $.

The importance of this conceptual subtility will appear as soon as
we will present the generalization of the Feynman-Kac formula to
arbitrary riemannian manifolds.
\newpage
\section{Anderson's nonstandard construction of the Wiener
measure on the euclidean line} \label{sec:Anderson's nonstandard
construction of the Wiener measure}

The powerful machinery of Loeb measures (reviewed in the section
\ref{sec:Loeb Probability Spaces}) has led R.M. Anderson
\cite{Anderson-76} to introduce a wonderful nonstandard
construction of the brownian motion  (as an infinitesimal random
walk) and of the Wiener measure.

Given the time interval $ ( 0 ,t ) , t \in ( 0 , + \infty) $ and
an unlimited hypernatural $ n \in \, ^{\star} \mathbb{N} -
\mathbb{N} $ let us introduce the  times:
\begin{equation}
    t_{k} \; := \;  k \epsilon \; \; k \in  \{ j \in \, ^{\star} \mathbb{N} : j \leq n \}
\end{equation}
where:
\begin{equation}
    \epsilon \; := \; \frac{ t }{n}
\end{equation}
Obviously:
\begin{equation}
    \epsilon \in hal(0)
\end{equation}
Let us consider the hyperfinite (time) interval:
\begin{equation}
    [ 0 , t]_{n} \; := \; \{  t_{k} \, , \,  k
\in  \{ j \in \, ^{\star} \mathbb{N} :  j \leq n \} \}
\end{equation}
and the set:
\begin{multline}
 \Omega ( t ; n ; q_{1} , q_{2} ) \; := \; \{ \omega : [0,t]
 \mapsto \, ^{\star} \mathbb{R} \; : \;  \omega (0) = q_{1} \, , \, \omega (t) =
 q_{2} \, , \\
 \omega ( t_{k+1} ) = \omega ( t_{k} ) \pm \sqrt{\epsilon} \, \text{ linearly interpolated between $ t_{k} $
and $ t_{k+1}$ } \\
  \forall k \in \{ j \in \, ^{\star} \mathbb{N}
:  j \leq n-1  \}  \}
\end{multline}

Let us now introduce the internal probability space $ (  \Omega (
t ; n ; q_{1} , q_{2} ) , \, ^{\star} \mathcal{P} [ \Omega ( t ; n
; q_{1} , q_{2} ) ] , W_{t ; n} ( q_{1} , q_{2} ) )  $, where $
W_{t ; n} ( q_{1} , q_{2} )  $ is the counting measure on $ \Omega
( t ; n ; q_{1} , q_{2} ) $ (see the section \ref{sec:Loeb
Probability Spaces}), and the corresponding Loeb probability space
$ (  \Omega ( t ; n ; q_{1} , q_{2} ) , L (^{\star} \mathcal{P} [
\Omega ( t ; n ; q_{1} , q_{2} ) ] ) , W_{t ; n} ( q_{1} , q_{2}
)_{L} ) $. Then Robert M. Anderson has proved the following:
\begin{theorem} \label{th:Anderson}
\end{theorem}
\emph{Anderson's Theorem:}
\begin{enumerate}
    \item for every Borel set $ B \subseteq \mathcal{C} ( q_{1}, 0 ; q_{2} , t
    ) $:
\begin{equation}
   W_{t ; n} (
q_{1} , q_{2} )_{L} ( st^{-1} (B) ) \; = \;   W_{t} ( q_{1} ,
q_{2} ) (B)
\end{equation}
where $ W_{t} ( q_{1} , q_{2} ) $ is the conditional Wiener
measure introduced in the previous section
    \item the stochastic process $ w : [ 0 , t ] \times \Omega ( t ; n ; q_{1} , q_{2}
    ) \mapsto \mathbb{R} $ defined by:
\begin{equation}
    w_{s} ( \omega ) \; := \; st [ \omega (s) ]
\end{equation}
is nothing but the brownian motion  over the metric space $ (
\mathbb{R} , d_{euclidean} ) $ introduced in the previous section.
\end{enumerate}

\smallskip

\begin{remark}
\end{remark}
Anderson's Theorem shows, in the particular case in which the
underlying riemannian manifold is the euclidean line, that
brownian motion can be defined as an infinitesimal random walk.

We will see that such a characterization can be extended to
Hausdorff riemannian manifolds.

\smallskip

\begin{remark}
\end{remark}

 Anderson's results have been  reformulated also in the language of  Internal Set
Theory \cite{Nelson-87}, \cite{Benoit-95}, \cite{Lawler-06}.

Since, as we show in section \ref{sec:Nonstandard Analysis going
outside ZFC: Internal Set Theory}, Internal Set Theory has great
conceptual bugs, such an approach, in our modest opinion, is
misleading.
\newpage
\section{Nonstandard Feynman-Kac formula on the euclidean line} \label{sec:Nonstandard Feynman-Kac formula on the euclidean line}

The idea of avoiding the mathematically meaningless continuum
limit of the  time-sliced path integrals by taking an hyperfinite
time-slicing has been already fruitfully adopted by many authors
\cite{Albeverio-Fenstad-Hoegh-Krohn-Lindstrom-86},
\cite{Albeverio-92}, \cite{Loo-99a}, \cite{Loo-99b},
\cite{Loo-00}.

Actually, it is sufficient to use the theorem \ref{th:Anderson} to
substitute Anderson's expression of the Wiener measure into the
Feynman-Kac formula on the euclidean real line (i.e. theorem
\ref{th:Feynman-Kac formula on the euclidean line}) to obtain the
following:
\begin{theorem} \label{th:Nonstandard  Feynman-Kac formula on the euclidean line:}
\end{theorem}
\emph{Nonstandard Feynman-Kac formula on the euclidean line:}
\begin{equation}
   K^{(V)}_{t}( q_{1}, q_{2} ) \; = \; \int_{\Omega ( t ; n ; q_{1} , q_{2} )} d [ W_{t ; n} (
q_{1} , q_{2} )_{L} ( st^{-1})]  \exp ( - \int_{0}^{t} d s V[q(s)]
   )
   \; \;  \forall q_{1} , q_{2} \in \mathbb{R} \, , \,  \forall  t \in ( 0 , +
  \infty ) \, , \, \forall n \in \, ^{\star} \mathbb{N} - \mathbb{N}
\end{equation}

\newpage
\section{The conditional Wiener measure and the Feynman-Kac formula on a riemannian manifold}

Given a D-dimensional riemannian manifold $ ( M , g) $ (with g
given in local coordinates by $ g = g_{\mu \nu} d x^{\mu} \otimes
d x^{\nu}$) let us consider the strongly-continuous contraction
semigroup $ \{ T^{(0)}_{t} := \exp ( - \frac{1}{2} t \triangle_{g}
) \}_{t \in ( 0 , + \infty )} $ of operators acting on the Hilbert
space $ \mathcal{H} := L^{2} ( M , d \mu_{g} ) $ where:
\begin{enumerate}
    \item  $ \triangle_{g} : \Omega^{r} (M ) \mapsto  \Omega^{r} (M
    ) \, , \, r \in \mathbb{N}$:
\begin{equation}
  \triangle_{g} \; := \;   d \, d^{\dag} +  d^{\dag} \, d
\end{equation}
(with $ \Omega^{r}(M) $ denoting the set of all r-forms on M) is
the Laplace-Beltrami operator on $ ( M , g ) $ \cite{Nakahara-03},
\cite{Frankel-04} of which we consider the restriction to $
\Omega^{0} (M) $ given in local coordinates by:
\begin{equation}
   \triangle_{g} f \; = \; - \frac{1}{\sqrt{| g | }}
   \partial_{\mu} [ \sqrt{| g |} g^{\mu \nu}  \partial_{\nu} f ]
\end{equation}
(where of course $ \partial_{\mu} := \frac{\partial}{\partial
x^{\mu} } $ while $ g  := det (g_{\mu \nu}) ) $
  \item $ d \mu_{g} $ is the invariant measure given in local
  coordinates by:
\begin{equation}
  d \mu_{g}  \; = \; \sqrt{| g |} d x^{1} \cdots d x^{D}
\end{equation}
\end{enumerate}

Introduced the kernel  $ K^{(0)}_{t}(
    q_{1}, q_{2} ) $ of $  \{ T^{(0)}_{t} \}_{t \in ( 0 , +
\infty)} $:
\begin{equation}
    ( T^{(0)}_{t} f ) (q_{1}) \; =: \int_{M} d \mu_{g}( q_{2}) \,  K^{(0)}_{t}(
    q_{1}, q_{2} ) f(q_{2}) \; \; \forall q_{1} \in
    M , \forall t \in ( 0 , + \infty )
\end{equation}
it may be proved  \cite{Fukushima-Oshima-Takeda-94} that $
K^{(0)}_{t} ( q_{1} , q_{2} ) $ satisfies the following
conditions:
\begin{enumerate}
    \item
\begin{equation}
   K^{(0)}_{t}( q_{1}, q_{2} ) > 0  \; \; \forall q_{1}, q_{2} \in
    M , \forall t \in ( 0 , + \infty )
\end{equation}
    \item
\begin{equation}
  \int_{M}  d \mu_{g}(q_{2}) \,  K^{(0)}_{t}( q_{1}, q_{2} ) \; = \; 1  \; \; \forall q_{1} \in
    M , \forall t \in ( 0 , + \infty )
\end{equation}
    \item
\begin{equation}
   K^{(0)}_{t_{1}+t_{2}}( q_{1}, q_{3} ) \; = \; \int_{M} d \mu_{g}(q_{2})
   \,  K^{(0)}_{t_{1}}( q_{1}, q_{2} )  K^{(0)}_{t_{2}}( q_{2}, q_{3} ) \; \; \forall q_{1}, q_{3} \in
    M , \forall t_{1} , t_{2} \in ( 0 , + \infty )
\end{equation}
\end{enumerate}
that guarantee that  $   K^{(0)}_{t}( q_{1}, q_{2} ) $ is the
transition probability kernel of a Markovian stochastic process $
w_{t} $ that one defines as the brownian motion  over  $ ( M , g)
$ \footnote{Such a definition may proved to be equivalent to the
one obtained through projection from the orthonormal frame bundle
\cite{Ikeda-Watanabe-81}, \cite{Elworthy-82}, \cite{Emery-89},
\cite{Taira-98}, \cite{Stroock-00}, \cite{Hsu-2002}.}.

Given $ q_{1} , q_{2} \in M $ and $ t \in ( 0 , + \infty ) $ let
us introduce the functional space:
\begin{equation}
  \mathcal{C} ( q_{1}, 0 ; q_{2} , t ) \; := \{ q : [ 0 , t ]
  \mapsto \mathbb{R} \; continuous \; : q(0) = q_{1} ,  q(t) =
  q_{2} \} \; \; q_{1} , q_{2} \in M , t \in ( 0 , +
  \infty )
\end{equation}
and the cylinder sets of $ \mathcal{C} ( q_{1}, 0 ; q_{2} , t ) $:
\begin{multline}
    \Gamma ( q_{1} , 0 ; \{ B_{k} ; t_{k} \}_{k=1}^{n-1} ;
    q_{2} , t ) \; := \; \{ q \in \mathcal{C} ( q_{1}, 0 ; q_{2} , t) : q(t_{k}) \in B_{k} \\
     \forall k \in \{ j \in \mathbb{N} \, : 1 \leq j \leq  n-1
    \} \} \; \; 0 < \cdots < t_{k} < t_{k+1}   < \cdots <  t ,  B_{k} \in
    \mathcal{B}_{Borel}( M )  \, \forall k \in \{ j \in \mathbb{N} \, : 1 \leq j \leq
    n-1 \} \, , \, n \in \mathbb{N}_{+}
\end{multline}
Defined the value of the conditional Wiener measure on cylinder
sets as:
\begin{equation}
    W [  \Gamma ( q_{1} , 0 ; \{ B_{k} ; t_{k} \}_{k=1}^{n-1} ;
    q_{2} , t ) ] \; := \; \prod_{k=1}^{n-1} \int_{ B_{k}} d \mu_{g} ( q_{k} ) \prod_{k=1}^{n}
    K^{(0)}_{t_{k+1}-t_{k}} ( q_{k} , q_{k+1} )
\end{equation}
it may be proved  that:
\begin{enumerate}
    \item the conditional Wiener measure $ W_{t} ( q_{1} , q_{2} )$ is countably-additive on
    the cylinder sets of $  \mathcal{C} ( q_{1}, 0 ; q_{2} , t ) $
    \item by Kolmogorov Reconstruction Theorem it has a unique
    extension to the Borel subsets of $  \mathcal{C} ( q_{1}, 0 ; q_{2} , t )
    $ that one defines to be the  conditional Wiener measure $ W_{t} ( q_{1} , q_{2} )$
\end{enumerate}

Given a potential energy V bounded from below, let us introduce
the one-parameter family of operators $ \{ T^{(V)}_{t} :=   \exp[
t ( - \frac{1}{2} \triangle_{g}  - V(q) )] \}_{t \in (0 , + \infty
)} $ and the associated integral kernel:
\begin{equation}
    ( T^{(V)}_{t} f ) (q_{1}) \; =: \int_{M} d \mu_{g}( q_{2}) \,  K^{(V)}_{t}(
    q_{1}, q_{2} ) f(q_{2}) \; \; \forall q_{1} \in
    M , \forall t \in ( 0 , + \infty )
\end{equation}

The application of the Trotter-Kato formula:
\begin{equation}
  \exp ( - \frac{1}{2} \triangle_{g}  - V(q) ) \; = \;
  s-\lim_{n \rightarrow + \infty} [ \exp ( - \frac{1}{2 n} \triangle_{g} ) \exp ( - \frac{V}{n} )]^{n}
\end{equation}
allows to derive the following (see the the section 11.4 "The
Feynman Integral and Feynman's Operational Calculus" of
\cite{Johnson-Lapidus-00} and the $ 7^{th}$ chapter "Symmetries"
of \cite{Cartier-De-Witt-Morette-06}):
\begin{theorem} \label{th:Feynman-Kac formula on a riemannian manifold}
\end{theorem}
\emph{Feynman-Kac formula on a riemannian manifold:}
\begin{equation}
   K^{(V)}_{t}( q_{1}, q_{2} ) \; = \; \int_{ \mathcal{C} ( q_{1}, 0 ; q_{2} , t
   )} d W_{t} ( q_{1} , q_{2} ) \exp ( - \int_{0}^{t} d s V[q(s)]
   )
   \; \;  \forall q_{1} , q_{2} \in M \, , \,  \forall  t \in ( 0 , +
  \infty )
\end{equation}

\begin{remark}
\end{remark}
Theorem \ref{th:Feynman-Kac formula on a riemannian manifold} is
strongly connected to a long-standing problem debated in the
Physics' literature, i.e. the one of quantizing a classical
non-relativistic dynamical system describing a particle of unary
mass constrained to move on $ ( M , g) $ having classical action
$S[q] := \int dt \frac{| \dot{q}|_{g}^{2} }{2} $ (see for instance
the $ 24^{th} $ chapter of \cite{Schulman-81}, the $ 9^{th} $
chapter "Quantization" of \cite{Woodhouse-94}, the $ 3^{th} $
chapter "Path integrals in Quantum Mechanics: Generalizations" of
\cite{Zinn-Justin-93}, the chapters 10 "Short-Time Amplitude in
Spaces with Curvature and Torsion" and the chapter 11
"Schr\"{o}dinger Equation in General Metric-Affine Spaces" of
\cite{Kleinert-95}, the $ 15^{th} $ chapter "The Nonrelativistic
Particle in a Curved Space" of \cite{De-Witt-03} as well as
references therein).

A great amount of the mentioned literature  is based on the
generalization of Feynman's approach, i.e. on:

\begin{enumerate}
    \item introducing $ n \in \mathbb{N}_{+} $  times:
\begin{equation}
    t_{k} \; := \;  k \epsilon \; \; k \in  \{ j \in \mathbb{N} \, : \, 0 \leq j \leq   n \}
\end{equation}
where:
\begin{equation}
    \epsilon \; := \; \frac{ t }{n}
\end{equation}
    \item using the semi-group property of the euclidean time-evolution
    operator $ \hat{U} $ in order to write:
\begin{equation}
  < q''  | \hat{U}_{t} |   q'  > \; = \; \prod_{k=1}^{n-1} \int_{M} d \mu_{g}( q_{k}) \prod_{k=1}^{n}  < q_{k}  | \hat{U}_{\epsilon} |   q_{k-1}  >
\end{equation}
where $ q_{0} := q' $ and $ q_{n} := q''$.
    \item obtaining, according to one among  different choices in the
    evaluation-point (claimed to correspond to different
    operators' orderings), one of the uncountably many possible expressions for $ < q_{k}  | \hat{U}_{\epsilon} |   q_{k-1}  >
    $ expressed in local coordinates.
    \item  performing a mathematically meaningless
    limit for $ n \rightarrow + \infty $ (and hence $ \epsilon
    \rightarrow 0 )$ in which the mathematically meaningless
    object $ \lim_{n \rightarrow + \infty}  ( \frac{1}{2 \pi \epsilon }
    )^{\frac{n}{2}} \prod_{k=1}^{n-1} d \mu_{g}(q_{k}) $ is replaced with
    a mathematically meaningless functional measure $ d \mu_{g} [ q(t) ]
    $
\end{enumerate}
whose mathematical meaning is less (if possible) than Feynman's
original approach owing to the  factor $ | g |^{\frac{n}{2}} $
obtained from the expressions in local coordinates of $ d \mu_{g}
$ of which each author gets rid of in some way.

\smallskip

Not surprisingly different authors arrive to different
conclusions.

\smallskip

Most of them concord on the fact that the quantum
    hamiltonian should be of the form $ \hat{H} \; = \;
    \frac{1}{2} \triangle_{g} + c R  $ where $ c \in \mathbb{Q} $
    and where R is the scalar curvature of $ ( M , g ) $.

    As to the value of the number c the more palatable proposals
    are $ c = 0$ (as according to Cecile Morette De-Witt), $
    c = \frac{1}{6}$ (in conformity with the $ 1^{th} $ order term in the
    asymptotic expansion $ K^{(0)}_{t} (q,q) \sim \sum_{n=0}^{+ \infty} a_{n}(q) t^{n} $ \cite{Gilkey-95}), $c = \frac{1}{12}$ (as
    according to the first Bryce De-Witt), $ c = \frac{1}{8} $
    (as according to the last Bryce De-Witt) and various other
    alternatives (someway related to the fact that, for $ D \geq 2 $, $\hat{H} $ is
    invariant under conformal transformations if and only if $ c =
    \frac{D-2}{4(D-1)}$).
\newpage
\section{Infinitesimal random walks on an Hausdorff riemannian manifold}
Nonstandard diffusions on manifolds have been studied in
\cite{Lindtsrom-86} using the machinery developed in the $ 5^{th}
$ chapter "Hyperfinite Dirichlet Forms and Markov Processes" of
\cite{Albeverio-Fenstad-Hoegh-Krohn-Lindstrom-86}.

The treatment therein contained, anyway, doesn't furnish an
explicit generalization of Anderson's construction.

The content of section  \ref{sec:Anderson's nonstandard
construction of the Wiener measure} may be easily generalized to
infinitesimal random walks over an Hausdorff riemannian manifold $
( M,g)$.

Given the time interval $ ( 0 ,t ) , t \in ( 0 , + \infty)  $ and
an unlimited hypernatural $ n \in \, ^{\star} \mathbb{N} -
\mathbb{N} $ let us introduce the times:
\begin{equation}
    t_{k} \; := \;  k \epsilon \; \; k \in  \{ j \in \, ^{\star} \mathbb{N} : j \leq n \}
\end{equation}
where:
\begin{equation}
    \epsilon \; := \; \frac{ t }{n}
\end{equation}
Obviously:
\begin{equation}
    \epsilon \in hal(0)
\end{equation}
Let us consider the hyperfinite (time) interval:
\begin{equation}
    [ 0 , t]_{n} \; := \; \{  t_{k} \,  k
\in  \{ j \in \, ^{\star} \mathbb{N} : j \leq n \} \}
\end{equation}

Given two points $ q_{1} , q_{2} \in M$:
\begin{definition}
\end{definition}
\begin{multline}
 \Omega ( t ; n ; q_{1} , q_{2} ) \; := \; \{ \omega : [ 0,t]
 \mapsto \, ^{\star} M \; : \;  \omega (0) = q_{1} \, , \, \omega (t) =
 q_{2} \, , \\  ^{\star} d_{g} [ \omega ( t_{k+1} ) , \omega ( t_{k} ) ]  = \pm
 \sqrt{\epsilon} \, , \\
  \text{ geodetically interpolated between $ t_{k}$ and $ t_{k+1} $ }  \forall k \in \{ j \in \, ^{\star} \mathbb{N} : 1 \leq j \leq n-1
 \} \}
\end{multline}
where the geodetic interpolation between $ \omega ( t_{k} ) $ and
$ \omega ( t_{k+1} ) $ is the extended shortest geodetic arc
connecting these points (whose existence is guaranteed by the
Hopf-Rinow Theorem \cite{Jost-95} combined with proposition
\ref{prop:transfer principle}) and $ ^{\star} d_{g} [ \omega (
t_{k+1} ) , \omega ( t_{k} ) ] $ is the extended geodesic-distance
between $ \omega ( t_{k} ) $ and $ \omega ( t_{k+1} ) $, i.e. the
length of such an infinitesimal geodetic arc.

Let us now introduce the internal probability space $ (  \Omega (
t ; n ; q_{1} , q_{2} ) , \, ^{\star} \mathcal{P} [ \Omega ( t ; n
; q_{1} , q_{2} ) ] , W_{t ; n} ( q_{1} , q_{2} ) )  $, where $
W_{t ; n} ( q_{1} , q_{2} )  $ is the counting measure on $ \Omega
( t ; n ; q_{1} , q_{2} ) $, and the corresponding Loeb
probability space $ (  \Omega ( t ; n ; q_{1} , q_{2} ) , L
(^{\star} \mathcal{P} [ \Omega ( t ; n ; q_{1} , q_{2} ) ] ) ,
W_{t ; n} ( q_{1} , q_{2} )_{L} ) $. Then:
\begin{theorem} \label{th:generalized Anderson's theorem}
\end{theorem}
\emph{Generalized Anderson's Theorem:}
\begin{enumerate}
    \item for every Borel set $ B \subseteq \mathcal{C} ( q_{1}, 0 ; q_{2} , t
    ) $:
\begin{equation}
   W_{t ; n} (
q_{1} , q_{2} )_{L} ( st^{-1} (B) ) \; = \;   W_{t} ( q_{1} ,
q_{2} ) (B)
\end{equation}
where $ W_{t} ( q_{1} , q_{2} ) $ is the conditional Wiener
measure  introduced in the previous section
    \item the stochastic process $ w : [ 0 , t ] \times \Omega ( t ; n ; q_{1} , q_{2}
    ) \mapsto \mathbb{R} $ defined by:
\begin{equation}
    w_{s} ( \omega ) \; := \; st [ \omega (s) ]
\end{equation}
is nothing but the brownian motion  over  $ ( M, g) $ introduced
in the previous section.
\end{enumerate}
\begin{proof}
 Anderson's proof can be completely formulated in terms of
Nonstandard Topology applied to the topological space $ (
\mathbb{R} , \mathcal{T}_{natural} ) $.

So, according to proposition \ref{prop:existence of the standard
part for Hausdorff topologies}, it may be immediately generalized
to M, seen as topological space, provided that it is Hausdorff.

The replacement of the euclidean mathematical objects of
Anderson's Theorem on the euclidean manifold $ ( \mathbb{R} ,
\delta ) $ with the corresponding riemannian-geometric objects of
the riemannian manifold $ ( M , g) $ is then straightforward.

\end{proof}
\smallskip
\section{Nonstandard Feynman-Kac formula on an Hausdorff riemannian manifold}

The approach followed in the section \ref{sec:Nonstandard
Feynman-Kac formula on the euclidean line} may be immediately
generalized to the case of an Hausdorff riemannian manifold $ ( M
, g )$.

Actually, it is sufficient to use the theorem \ref{th:generalized
Anderson's theorem} to substitute the generalized Anderson's
expression of the Wiener measure into the Feynman-Kac formula on
$(M,g)$ (i.e. the theorem \ref{th:Feynman-Kac formula on a
riemannian manifold}) to obtain the following:
\begin{theorem} \label{th:Nonstandard  Feynman-Kac formula on an Hausdorff riemannian manifold}
\end{theorem}
\emph{Nonstandard Feynman-Kac formula on an Hausdorff riemannian
manifold:}
\begin{equation}
   K^{(V)}_{t}( q_{1}, q_{2} ) \; = \; \int_{\Omega ( t ; n ; q_{1} , q_{2} )} d [ W_{t ; n} (
q_{1} , q_{2} )_{L} ( st^{-1})]  \exp ( - \int_{0}^{t} d s V[q(s)]
   )
   \; \;  \forall q_{1} , q_{2} \in M \, , \,  \forall  t \in ( 0 , +
  \infty ) \, , \, \forall n \in \, ^{\star} \mathbb{N} - \mathbb{N}
\end{equation}
\newpage
\section{Taking into account noncommutating operators: infinitesimal quantum random walks on noncommutative riemannian manifolds}
As the link between Quantum Mechanics and brownian motion
exhibited by the Feynman-Kac formulas led Edward Nelson to
formulate Stochastic Mechanics \cite{Nelson-67}, \cite{Nelson-85},
a reformulation of Quantum Mechanics in terms of classical
markovian stochastic processes in which, for instance, the
quantization of the nonrelativistic classical dynamical system
consisting in a particle of unary mass constrained to the
riemannian manifold $ ( M, g) $ (and hence described by the
classical action functional $ S[q] := \int dt \frac{|
\dot{q}|_{g}^{2} }{2} $) is performed entirely in terms of
stochastic averages with respect to the brownian motion on $ (
M,g) $, one could think to adopt theorem \ref{th:Nonstandard
Feynman-Kac formula on an Hausdorff riemannian manifold} as a
starting point to formulate a nonstandard version of Stochastic
Mechanics in which, for instance, the quantization of the
mentioned classical dynamical system would be performed entirely
in terms of stochastic averages with respect to the infinitesimal
random walk on $ ( M,g) $.

From a mathematical point of view this would allow to formulate in
a mathematically and conceptually rigorous way the stochastic
variational approach bypassing the problem of the ill-defined
nature of the stochastic functional $ S_{stoc} [q] := \int dt E[
\frac{| \dot{q}|_{g}^{2} }{2} ]$ (owed to the basic property of
brownians paths informally expressed as $ (d q)^{2} \, = \, dt $
 \, \footnote{a consequence of which is the fact that the Wiener
measure is supported on paths H\"{o}lderian of order $ <
\frac{1}{2} $ or, in a more fashionable language, that brownian
paths have Hausdorff dimension 2.} and the consequential informal
fact that $ \dot{q} :=  \sqrt{\frac{(dq)^{2}}{(dt)^{2}} } \; = \;
\sqrt{\frac{1}{dt}} $ is ill-defined) and hence also the
"resolution" of the problem through the adoption of the
controversial renormalization of the stochastic action that, in
our modest opinion, is no more than a conjuring trick.

A consistent stochastic variational principle would then be
formalizable in terms of the Loeb Measure Theoretic approach to
the  Malliavin Calculus \cite{Nualart-95}, \cite{Malliavin-97}
exposed in the $ 3^{th} $ chapter "Stochastic Calculus of
Variations" of \cite{Cutland-00}.

\smallskip

This would, anyway, contribute to hide a structural limitation of
the Feynman-Kac formulas (and consequentially of Stochastic
Mechanics) that it is never sufficiently remarked:

though they allow to express the quantum averages of the elements
of the commutative Von Neumann subalgebra generated by one
operator as expectation values taken with respect to a classical
(i.e. commutative) Kolmogorovian probability space, the game
breaks up as soon as one takes into accounts noncommutating
operators (whose existence is the soul of Quantum Mechanics), as
it can be appreciated going higher than the ground floor of the
following (see for instance \cite{Kadison-Ringrose-97a},
\cite{Kadison-Ringrose-97b}, \cite{Parthasarathy-92},
\cite{Cuculescu-Oprea-94} \cite{Meyer-95}, \cite{Manin-91},
\cite{Connes-94}, \cite{Landi-97}, \cite{Connes-98},
\cite{Gracia-Bondia-Varilly-Figueroa-01}):
\begin{theorem} \label{th:noncommutative tower}
\end{theorem}
\emph{Theorem of the Noncommutative Tower:}
\begin{itemize}
    \item (ground floor) Noncommutative Topology:
    \begin{center}
The \emph{category} having as \emph{objects} the \emph{Hausdorff
compact topological spaces} and as \emph{morphisms} the
\emph{continuous maps} on such spaces is equivalent to the
category having as \emph{objects} the \emph{abelian
$C^{\star}$-algebras} and as \emph{morphisms}  the
\emph{involutive morphisms} of such spaces.
    \end{center}
    \item (first floor) Noncommutative Probability:
    \begin{center}
    The \emph{category} having as \emph{objects} the
\emph{classical probability spaces} and as \emph{morphisms} the
\emph{endomorphisms (automorphisms)} of such spaces is equivalent
to the \emph{category} having as objects the \emph{abelian
algebraic probability  spaces} and as \emph{ morphisms}  the
\emph{endomorphisms (automorphisms)} of such spaces.
    \end{center}
    \item (second floor) Noncommutative Geometry:
    \begin{center}
The \emph{category} having as \emph{objects} the \emph{closed
finite-dimensional riemannian spin manifolds} and as
\emph{morphisms} the \emph{diffeomorphisms of such manifolds} is
equivalent to the \emph{category} having as \emph{objects} the
\emph{abelian spectral triples} and as \emph{morphisms} the
\emph{automorphisms of the involved involutive algebras}.
    \end{center}

\end{itemize}
to infer that as a matter of principle only a limited number of
moments of a noncommutative probability space can be reproduced by
a commutative probability space (as concretely shown by the
difficulties arising as soon as one tries to give a mathematical
foundation to Feynman's operational calculus in terms of
integration with respect to the Wiener measure
\cite{Johnson-Lapidus-00}) \footnote{One could, at this point,
object that to make such an inference it is enough to go to the
first floor and there is no necessity to raise to the second
floor. This leads us directly do the issue about the hierarchy
existing between Probability and Geometry. The $ \sigma$-algebra
of the measurable spaces most used in Theoretical and Mathematical
Physics is the Borel-$\sigma$ algebra induced by some topology. In
many cases such a topology is the (Hausdorff) metric one induces
by a metric depending from an  underlying geometric structure; for
example it could be the geodesic distance arising from a
riemannian manifold's structure. As to the set of all the
probability measures on a measurable space, Information Geometry
\cite{Amari-Kagaoka-00} taught us its underlying geometric
structure. Hence, despite the appearances, in many cases we see
that Geometry is hierarchically prior than Measure Theory (and
hence Probability Theory). The same situation occurs in the
noncommutative context where in many cases the Von Neumann algebra
A of a noncommutative probability space $ ( A , \omega) $ has a
noncommutative geometric underlying structure (unfortunately
appreciable only after having digested  $ C^{\star}$-modules,
cyclic cohomology, crossed products and whatsoever).}.

Despite Connes' criticisms concerning the claimed nonconstructive
nature of the infinitesimals of Nonstandard Analysis (about which
we demand to the remark \ref{rem:hyperreals and constructivism})
and his interpretation of the compact operators as the right
infinitesimals in the noncommutative framework, it should be
possible to use Nonstandard Analysis to reformulate the
Sinha-Goswami's definition of a quantum brownian motion on a
noncommutative riemannian manifold as an Evans-Hudson dilation of
the heat semigroup of the underlying $ C^{\star}$-algebra (see the
$ 9^{th} $ chapter "Noncommutative Geometry and quantum stochastic
process " of \cite{Sinha-Goswami-07}) looking at such a quantum
brownian motion as an infinitesimal quantum random walks on such a
noncommutative riemannian manifold by making the usual ansatz in
which  a time interval $ [ 0,t] $, where $ t \in ( 0 , + \infty )
$, is replaced with an hyperfinite time interval $ [ 0,t]_{n} \, ,
\, n \in \, ^{\star} \mathbb{N} - \mathbb{N} $.

A preliminary task in this direction would consist in recovering
Anderson's construction from the representation of classical
brownian motion on $ ( \mathbb{R} , d_{euclidean} ) $ in terms of
(what it is natural to call) the hyperfinite family of operators:
\begin{equation}
    \{ ( \hat{a}+ \hat{a}^{\dag})(s)  \; s \in [ 0 , t]_{n} \}
\end{equation}
on the symmetric Fock space $ \Gamma [ L^{2} ( [0, + \infty) , dq)
\otimes \mathbb{C} ] $ where $ \hat{a} $ and $ \hat{a}^{\dag} $
are the usual, respectively, annihilation and creation operators:
\begin{equation}
    [ \hat{a} (s) , \hat{a}^{\dag} (s) ] \; = \; 1 \; \; \forall s
    \in  [ 0 , t]_{n}
\end{equation}
and where as usual $ [ 0 , t ]_{n} $, for $ t \in ( 0 , + \infty )
$ and $ n \in \, ^{\star} \mathbb{N} - \mathbb{N} $, is the
hyperfinite (time) interval:
\begin{equation}
    [ 0 , t ]_{n} \; := \; \{  k \cdot \epsilon \, , \,  k
\in  \{ j \in \, ^{\star} \mathbb{N} : j \leq n \} \}
\end{equation}
\begin{equation}
    \epsilon \; := \; \frac{ t }{n} \; \in \; hal(0)
\end{equation}

\newpage
\section{Acknowledgements}
I would like to thank  Vittorio de Alfaro for his friendly support
and Piergiorgio Odifreddi for his real help about surreal numbers.
\newpage
\appendix
\section{The orthodox ZFC+CH set-theoretic foundation of
Mathematics} \label{sec:The orthodox ZFC+CH set-theoretic
foundation of Mathematics}

It is nowadays common opinion in the Scientific Community that:
\begin{enumerate}
    \item The foundations of Mathematics lies on Set Theory
    \item Set Theory is axiomatized by the formal system ZFC, i.e. the
Zermelo-Fraenkel formal system augmented with the Axiom of Choice
\end{enumerate}
In this section we will strongly defend such an orthodox viewpoint
presenting ZFC in detail \cite{Ciesielski-97}.

Before the birth of ZFC Set Theory was studied by a naive approach
(that we will denote as Naive Set Theory)  based on the following:
\begin{axiom} \label{ax:Frege's comprehension}
\end{axiom}
\emph{Axiom of Frege's-Comprehension:}
\begin{center}
    If $ p $ is a unary predicate then there exists a set $ \mathbb{U}_{p}
    := \{ x : p(x) \} $ of all elements having the property p
\end{center}
Naife Set Theory was proved to be inconsistent by Russell who
showed that the application of the axiom \ref{ax:Frege's
comprehension} to the unary predicate $ p_{Russell}( x) := x
\notin x $ leads to the contradiction:
\begin{equation}
    S_{p_{Russell}} \in  S_{p_{Russell}} \; \Leftrightarrow \; S_{p_{Russell}} \notin  S_{p_{Russell}}
\end{equation}

The conceptual earthquake caused by Russell's remark led the
Scientific Community to think that a more refined axiomatization
of Set Theory was needed.

The result of the deep work of many mathematicians was the
formulation of the following:
\begin{definition} \label{def:ZF}
\end{definition}
\emph{Formal System of Zermelo-Fraenkel (ZF):}
\begin{enumerate}
    \item
\begin{axiom} \label{ax:existence of the empty set}
\end{axiom}
\emph{Axiom of Existence of the empty set:}

There exist the empty set:
\begin{equation}
    \exists \emptyset \; : \; \forall x  \; \neg ( x \in \emptyset
    )
\end{equation}
    \item
\begin{axiom}
\end{axiom}
\emph{Axiom of Extensionality:}

If x and y have the same elements, then x is equal to y:
\begin{equation}
    \forall x \forall y \, [ \forall z \, ( z \in x
  \,  \Leftrightarrow \, z \in y ) \, \Rightarrow \, x= y ]
\end{equation}
    \item
\begin{axiom} \label{ax:comprehension}
\end{axiom}
\emph{Axiom of Comprehension:}

For every formula $ \phi (s,t) $ with free variables s and t, for
every x, and for every parameter p there exists a set $ y : = \{ u
\in x : \phi(u,p) \} $:
\begin{equation}
    \forall x \forall p \exists y \; : \; [ \forall u ( u \in y
    \, \Leftrightarrow \, ( u \in x  \wedge \phi (u,p)))]
\end{equation}
    \item
\begin{axiom}
\end{axiom}
\emph{Axiom of Pairing:}

For any a and b there exists a set x that contains a and b:
\begin{equation}
    \forall a \forall b \, \exists x \, : ( a \in x \, \wedge b
    \in x )
\end{equation}
    \item
\begin{axiom}
\end{axiom}
\emph{Axiom of Union:}

For every family $ \mathcal{F} $ there exists a set U containing
the union $ \cup \mathcal{F} $ of all the elements of $
\mathcal{F} $:
\begin{equation}
    \forall \mathcal{F} \, \exists U : \forall Y \forall x [ x \in
    Y \, \wedge \, Y \in \mathcal{F} ) \Rightarrow x \in U]
\end{equation}
    \item
\begin{axiom}
\end{axiom}
\emph{Axiom of the Power Set:}

For every set X there exists a set P containing the set $
\mathcal{P} (X) $ (called the power set of X) of all subsets of X:
\begin{equation}
    \forall X \exists P \, : \: \forall z [ z \subset X
    \Rightarrow z \in P ]
\end{equation}
    \item
\begin{axiom} \label{ax:infinity}
\end{axiom}
\emph{Axiom of Infinity:}

there exists an infinite set (of some special form):
\begin{equation}
  \exists x : [ \forall z ( z = \emptyset \Rightarrow z \in x ) \:
  \wedge \: \forall y \in x \forall z ( z= Suc(y) \Rightarrow z \in x
  )]
\end{equation}
where $ Suc(x) := x \cup \{ x \} $ is called the successor of x.
    \item
\begin{axiom}
\end{axiom}
\emph{Axiom of Replacement:}

For every formula $ \phi (s,t,U,w) $ with free variables s,t,U and
w, for every set A and for every parameter p, if $ \phi (s,t,A,p)
$ defines a function F on A by $ F(x) = y  \Leftrightarrow \phi
(x,y,A,p) $, then there exists a set Y containing the range $ F[A]
:= \{ F(x) : x \in A \} $ of the function F:
\begin{equation}
    \forall A \forall p [ \forall x \in A \, \exists ! y : \phi
(x,y,A,p) \; \Rightarrow \; \exists Y : \forall x \in A  \exists y
\in Y : \phi (x,y,A,p)]
\end{equation}
\item
\begin{axiom} \label{ax:foundation}
\end{axiom}
\emph{Axiom of Foundation:}

Every non-empty set has an $ \in $ -minimal element:
\begin{equation}
    \forall x [ \exists y : ( y \in x) \Rightarrow \exists y : ( y
    \in x \, \wedge \, \neg \exists z : ( z \in x \wedge z \in y
    ))]
\end{equation}

\end{enumerate}

\bigskip

\begin{remark}
\end{remark}
Let us remark that, to shorten the notation, we have used in the
definition \ref{def:ZF} the symbol $ \cup $ of union and the
symbol $ \subset $ of inclusion though the only (undefined) unary
predicate contained in the definition \ref{def:ZF} is the
predicate $ \in $ of memberships.

Actually all the other set-theoretic connectives are defined in
terms of it.

So, given two sets $ S_{1} $ and $ S_{2} $:
\begin{definition}
\end{definition}
\emph{union of $ S_{1} $ and $ S_{2}$:}
\begin{equation}
    S_{1} \cup S_{2} \; := \{ x : x \in S_{1} \; \vee \; x \in S_{2} \}
\end{equation}
\begin{definition}
\end{definition}
\emph{intersection of $ S_{1} $ and $ S_{2}$}
\begin{equation}
    S_{1} \cap S_{2} \; := \{ x : x \in S_{1} \; \wedge \; x \in S_{2} \}
\end{equation}
\newpage
\begin{definition}
\end{definition}
\emph{$ S_{1} $ is a subset of $ S_{2}$:}
\begin{equation}
    S_{1} \subseteq S_{2} \; := \; x \in S_{2} \; \; \forall x \in
    S_{1}
\end{equation}
\begin{definition}
\end{definition}
\emph{$ S_{1} $ is a proper subset of $ S_{2}$:}
\begin{equation}
    S_{1} \subset S_{2} \; := \;  S_{1} \subseteq S_{2} \: \wedge
    \: S_{1} \neq S_{2}
\end{equation}

\smallskip

\begin{remark} \label{rem:nonconstructive nature of the Axiom of Foundation}
\end{remark}
Let us remark that the Axiom of Foundation (i.e. the axiom
\ref{ax:foundation}) is nonconstructive: it assures us that given
a non-empty set S there exist an element m(S) of S that is $ \in
$-minimal, but it doesn't gives an algorithm that, receiving S as
input, gives m(S) as output.

\smallskip

Let us assume the formal system ZF.

\smallskip

Given arbitrary a and b:
\begin{definition}
\end{definition}
\emph{ordered pair of a and b:}
\begin{equation}
    < a , b > \; := \; \{ \{ a \} , \{ a , b \} \}
\end{equation}

Given two sets $ S_{1} $ and $ S_{2} $:
\begin{definition}
\end{definition}
\emph{cartesian product of $ S_{1} $ and $ S_{2}$:}
\begin{equation}
    S_{1} \times S_{2} \; := \; \{ z \in \mathcal{P} (\mathcal{P}
    ( S_{1} \cup S_{2} )) \, : \,  \exists x \in S_{1} \exists y \in
    S_{2} : ( z = < x , y > ) \}
\end{equation}

\begin{definition}
\end{definition}
\emph{binary relation between $ S_{1} $ and $ S_{2}$:}
\begin{equation}
    R \in \mathcal{P} ( S_{1} \times S_{2} )
\end{equation}
Given a binary relation R between $ S_{1} $ and $ S_{2} $ let
introduce the notation:
\begin{definition}
\end{definition}
\begin{equation}
    x R y \; := \; < x , y > \in R
\end{equation}
\begin{definition}
\end{definition}
\emph{R is a map with domain $S_{1}$ and codomain $ S_{2}$:}
\begin{equation}
    \forall x \in S_{1} , \exists ! R(x) \in S_{2} \; : \, x R
    R(x)
\end{equation}
A map f with domain $S_{1}$ and codomain $ S_{2}$ (briefly a map
from $ S_{1} $ to $ S_{2} $) is denoted as $ f : S_{1} \mapsto
S_{2} $.
\begin{definition}
\end{definition}
\emph{set of all maps with domain $S_{1}$ and codomain $ S_{2}$ }
\begin{equation}
    S_{2}^{S_{1}} \; := \; \{ f : S_{1} \mapsto S_{2}   \}
\end{equation}

Given a binary relation R on a set S:
\begin{definition}
\end{definition}
\emph{R is a preordering:}

\begin{enumerate}
    \item it is reflexive:
\begin{equation}
    x R x \; \; \forall x \in S
\end{equation}
    \item it is transitive:
\begin{equation}
   [ ( x_{1} R x_{2} \, \wedge \,  x_{2} R x_{3} ) \; \Rightarrow
    \;  x_{1} R x_{3} ] \; \; \forall x_{1},x_{2},x_{3} \in S
\end{equation}
\end{enumerate}

\begin{definition}
\end{definition}
\emph{R is an equivalence relation:}

\begin{enumerate}
    \item it is a preordering
    \item it is symmetric:
\begin{equation}
  ( x_{1} R x_{2} \; \Rightarrow \;  x_{2} R x_{1} ) \; \; \forall
  x_{1}, x_{2} \in S
\end{equation}
\end{enumerate}

\begin{definition}
\end{definition}
\emph{R is a partial ordering over S:}
\begin{enumerate}
    \item it is a preordering
    \item it is antisymmetric:
\begin{equation}
  [ ( x_{1} R x_{2} \wedge  x_{2} R x_{1} ) \; \Rightarrow \; x_{1}
    = x_{2} ] \; \; \forall x_{1}, x_{2} \in S
\end{equation}
\end{enumerate}

\begin{definition}
\end{definition}
\emph{R is a total ordering over S:}
\begin{enumerate}
    \item it is a partial ordering over S
    \item
\begin{equation}
  ( x_{1} R x_{2} \; \vee \;  x_{2} R x_{1} ) \; \; \forall
  x_{1}, x_{2} \in S
\end{equation}
\end{enumerate}

\begin{definition}
\end{definition}
\emph{R is well-founded:}
\begin{equation}
    [ Y \neq \emptyset \, \Rightarrow \, \exists m \in Y :
   ( \nexists y \in Y : y R m ) ] \; \; \forall Y
   \subset S
\end{equation}

\begin{definition}
\end{definition}
\emph{R is a well-ordering over S}:
\begin{enumerate}
    \item it is a total ordering over S
    \item it is well-founded
\end{enumerate}

\bigskip

We have now all the ingredients required to introduce in a compact
way the following:
\newpage
\begin{definition} \label{def:ZFC}
\end{definition}
\emph{formal system of Zermelo-Fraenkel augmented with the Axiom
of Choice (ZFC):}

the formal system ZF augmented with the following:
\begin{axiom} \label{ax:choice}
\end{axiom}
\emph{Axiom of Choice:}
\begin{equation}
   \forall S \neq \emptyset \;  \exists f \in (\cup_{B \in \mathcal{P}
    (S)} B)^{ \mathcal{P} (S)} \; : \; f(A) \in A \; \; \forall A \in  \mathcal{P} (S) :
    A \neq \emptyset
\end{equation}

\smallskip

\begin{remark} \label{rem:nonconstructive nature of the Axiom of Choice}
\end{remark}
Let us remark that the Axiom of Choice (i.e. axiom
\ref{ax:choice}) is nonconstructive: it assures us that given a
non-empty set S there exists a function $ f_{S} $ that maps each
non-empty subset into an element of it but it doesn't gives us an
algorithm that, receiving as input the set S, gives as output the
map $ f_{S} $.

\begin{remark}
\end{remark}
The nonconstructive nature of the Axiom of Choice (i.e. axiom
\ref{ax:choice}) has led many mathematicians and physicists to
look with suspicion at the results depending on it (such as the
Hahn-Banach Theorem in Functional Analysis).

G\"{o}del has proved that the Axiom of Choice  is consistent
relative to ZF, i.e. that if ZF is consistent then ZF augmented
with the Axiom of Choice is consistent too.

However Paul Cohen has proved that also the negation of the Axiom
of Choice is consistent relative to ZF.

Horst Herrlich has recently published a very interesting book
\cite{Herrlich-06} comparing the disasters occurring avoiding the
Axiom of Choice with the ones occurring assuming the Axiom of
Choice.

As to Mathematical-Physics it should be remarked that while the
disasters (such as the famous Banach-Tarski Paradox according to
which, within ZFC, any two bounded subsets A and B of $
\mathbb{R}^{3}$, each one containing some ball, are
equidecomposable) caused by the Axiom of Choice are easily
exorcizable (as to the Banach-Tarski Paradox, for instance, this
is automatically done by the non Lebesgue-measurability of the
pieces of the paradoxical decompositions) the disasters caused by
the absence of the Axiom of Choice (such as vector spaces having
no bases or having bases of different cardinalities) drastically
compromise any mathematical foundation of Quantum Mechanics.

\smallskip

Let us  assume the formal system ZFC of which we want here to show
some key features:

\begin{theorem}
\end{theorem}
\emph{Zermelo's  Theorem:}
\begin{equation}
    \forall S \neq \emptyset \; \exists  R : R \text{ is a well-ordering over S}
\end{equation}

\smallskip

\begin{remark}
\end{remark}
Let us remark that the Axiom of Comprehension (i.e. axiom
\ref{ax:comprehension}) is different from the Axiom of
Frege's-Comprehension (i.e. axiom \ref{ax:Frege's comprehension})
since given a unary predicate p:
\begin{enumerate}
    \item given a set S it allows to
    define the set $ S_{p} := \{ x \in S : p(x) \} $
    \item  since the undefined object "the proper class $ \mathbb{U} $ of all sets" is not a set in ZFC
it doesn't exist within
    ZFC a set $ \mathbb{U}_{p} := \{ x \in \mathbb{U} : p(x) \} $
\end{enumerate}
and hence Russell's paradox doesn't occur within ZFC.

\smallskip

\begin{remark}
\end{remark}
The axiom \ref{ax:foundation} (that can be compactly stated as the
condition that the binary relation $ \in $ is well-founded on
every non-empty set) implies that:
\newpage
\begin{theorem}
\end{theorem}
\begin{enumerate}
    \item
\begin{equation}
    S \notin S \; \; \forall S
\end{equation}
    \item
\begin{equation}
  \nexists \{ S_{n} \}_{n \in \mathbb{N}} \; : \; ( S_{n+1} \in
  S_{n} \; \; \forall n \in \mathbb{N} )
\end{equation}
\end{enumerate}

\smallskip
\begin{remark}
\end{remark}
The Axiom of Existence of the Empty Set (i.e. axiom
\ref{ax:existence of the empty set}) together with the Axiom of
Infinity (i.e. axiom \ref{ax:infinity}) allows to prove the
following:
\begin{theorem} \label{eq:existence  and unicity of ZFC's naturals}
\end{theorem}
\emph{Existence and unicity of the set of all natural numbers:}
\begin{center}
 There exists exactly one set $  \mathbb{N} $ such that:
\begin{enumerate}
    \item
\begin{equation} \label{eq:initial condition}
    \emptyset \in \mathbb{N}
\end{equation}
    \item
\begin{equation}  \label{eq:induction condition}
    Suc(x) \in \mathbb{N} \; \; \forall x \in \mathbb{N}
\end{equation}
where, as in the Axiom of Infinity (i.e axiom \ref{ax:infinity}),
$ Suc(x) := x \cup \{ x \} $.
    \item if K is any set that satisfies eq. \ref{eq:initial
    condition} and eq. \ref{eq:induction condition}  then $
    \mathbb{N} \subset K$.
\end{enumerate}
\end{center}
\begin{proof}
By the Axiom of Infinity (i.e. axiom \ref{ax:infinity}) there
exists at least one set X satisfying eq. \ref{eq:initial
condition} and eq. \ref{eq:induction condition}.

Let:
\begin{equation}
    \mathcal{F} \; := \; \{ Y \in \mathcal{P} (X) : \emptyset \in Y \, \wedge \, ( Suc(x) \in Y \; \forall x \in Y ) \}
\end{equation}
\begin{equation}
    \mathbb{N} \; := \; \bigcap_{Y \in  \mathcal{F} } Y
\end{equation}
It is easy to see that the intersection of any nonempty family of
sets satisfying  eq. \ref{eq:initial condition} and eq.
\ref{eq:induction condition} still satisfies  eq. \ref{eq:initial
condition} and eq. \ref{eq:induction condition}.

Let K be a set that satisfies  eq. \ref{eq:initial condition} and
eq. \ref{eq:induction condition}; then:
\begin{equation}
    X \cap K \in \mathcal{F}
\end{equation}
and:
\begin{equation}
     \mathbb{N} \; := \; \bigcap_{Y \in  \mathcal{F} } Y \;
     \subset X \cap K \; \subset \; K
\end{equation}
\end{proof}

\begin{remark}
\end{remark}

Let us remark that the set $ \mathbb{N} $ whose existence and
unicity is stated by theorem \ref{eq:existence and unicity of
ZFC's naturals} is  recursive \cite{Cutland-80},
\cite{Odifreddi-89}.

Actually every element of its can be concretely computed through
the following Mathematica \cite{Wolfram-96} expressions:
\begin{verbatim}
$RecursionLimit=Infinity;

 natural[n_] := If[n == 0, {}, Union[natural[n - 1], {natural[n - 1]}]]
\end{verbatim}
(where a finite set is represented by a list and where the empty
set $ \emptyset $ is represented by the empty list $ \{ \} $) from
which we obtain, for instance, that:
\begin{verbatim}
0 = {}

1 = {{}}

2 = {{},{{}}}

3 = {{},{{}},{{},{{}}}}

4 = {{},{{}},{{},{{}}},{{},{{}},{{},{{}}}}}

5 =
{{},{{}},{{},{{}}},{{},{{}},{{},{{}}}},{{},{{}},{{},{{}}},{{},{{}},{{},{{}}}}}}

6 =
{{},{{}},{{},{{}}},{{},{{}},{{},{{}}}},{{},{{}},{{},{{}}},{{},{{}},{{},{{}}}}}\
,{{},{{}},{{},{{}}},{{},{{}},{{},{{}}}},{{},{{}},{{},{{}}},{{},{{}},{{},{{}}}}\
}}}


7 =
{{},{{}},{{},{{}}},{{},{{}},{{},{{}}}},{{},{{}},{{},{{}}},{{},{{}},{{},{{}}}}}\
,{{},{{}},{{},{{}}},{{},{{}},{{},{{}}}},{{},{{}},{{},{{}}},{{},{{}},{{},{{}}}}\
}},{{},{{}},{{},{{}}},{{},{{}},{{},{{}}}},{{},{{}},{{},{{}}},{{},{{}},{{},{{}}\
}}},{{},{{}},{{},{{}}},{{},{{}},{{},{{}}}},{{},{{}},{{},{{}}},{{},{{}},{{},{{}\
}}}}}}}

8 =
{{},{{}},{{},{{}}},{{},{{}},{{},{{}}}},{{},{{}},{{},{{}}},{{},{{}},{{},{{}}}}}\
,{{},{{}},{{},{{}}},{{},{{}},{{},{{}}}},{{},{{}},{{},{{}}},{{},{{}},{{},{{}}}}\
}},{{},{{}},{{},{{}}},{{},{{}},{{},{{}}}},{{},{{}},{{},{{}}},{{},{{}},{{},{{}}\
}}},{{},{{}},{{},{{}}},{{},{{}},{{},{{}}}},{{},{{}},{{},{{}}},{{},{{}},{{},{{}\
}}}}}},{{},{{}},{{},{{}}},{{},{{}},{{},{{}}}},{{},{{}},{{},{{}}},{{},{{}},{{},\
{{}}}}},{{},{{}},{{},{{}}},{{},{{}},{{},{{}}}},{{},{{}},{{},{{}}},{{},{{}},{{}\
,{{}}}}}},{{},{{}},{{},{{}}},{{},{{}},{{},{{}}}},{{},{{}},{{},{{}}},{{},{{}},{\
{},{{}}}}},{{},{{}},{{},{{}}},{{},{{}},{{},{{}}}},{{},{{}},{{},{{}}},{{},{{}},\
{{},{{}}}}}}}}}

9 =
{{},{{}},{{},{{}}},{{},{{}},{{},{{}}}},{{},{{}},{{},{{}}},{{},{{}},{{},{{}}}}}\
,{{},{{}},{{},{{}}},{{},{{}},{{},{{}}}},{{},{{}},{{},{{}}},{{},{{}},{{},{{}}}}\
}},{{},{{}},{{},{{}}},{{},{{}},{{},{{}}}},{{},{{}},{{},{{}}},{{},{{}},{{},{{}}\
}}},{{},{{}},{{},{{}}},{{},{{}},{{},{{}}}},{{},{{}},{{},{{}}},{{},{{}},{{},{{}\
}}}}}},{{},{{}},{{},{{}}},{{},{{}},{{},{{}}}},{{},{{}},{{},{{}}},{{},{{}},{{},\
{{}}}}},{{},{{}},{{},{{}}},{{},{{}},{{},{{}}}},{{},{{}},{{},{{}}},{{},{{}},{{}\
,{{}}}}}},{{},{{}},{{},{{}}},{{},{{}},{{},{{}}}},{{},{{}},{{},{{}}},{{},{{}},{\
{},{{}}}}},{{},{{}},{{},{{}}},{{},{{}},{{},{{}}}},{{},{{}},{{},{{}}},{{},{{}},\
{{},{{}}}}}}}},{{},{{}},{{},{{}}},{{},{{}},{{},{{}}}},{{},{{}},{{},{{}}},{{},{\
{}},{{},{{}}}}},{{},{{}},{{},{{}}},{{},{{}},{{},{{}}}},{{},{{}},{{},{{}}},{{},\
{{}},{{},{{}}}}}},{{},{{}},{{},{{}}},{{},{{}},{{},{{}}}},{{},{{}},{{},{{}}},{{\
},{{}},{{},{{}}}}},{{},{{}},{{},{{}}},{{},{{}},{{},{{}}}},{{},{{}},{{},{{}}},{\
{},{{}},{{},{{}}}}}}},{{},{{}},{{},{{}}},{{},{{}},{{},{{}}}},{{},{{}},{{},{{}}\
},{{},{{}},{{},{{}}}}},{{},{{}},{{},{{}}},{{},{{}},{{},{{}}}},{{},{{}},{{},{{}\
}},{{},{{}},{{},{{}}}}}},{{},{{}},{{},{{}}},{{},{{}},{{},{{}}}},{{},{{}},{{},{\
{}}},{{},{{}},{{},{{}}}}},{{},{{}},{{},{{}}},{{},{{}},{{},{{}}}},{{},{{}},{{},\
{{}}},{{},{{}},{{},{{}}}}}}}}}}

10 =
{{},{{}},{{},{{}}},{{},{{}},{{},{{}}}},{{},{{}},{{},{{}}},{{},{{}},{{},{{}}}}}\
,{{},{{}},{{},{{}}},{{},{{}},{{},{{}}}},{{},{{}},{{},{{}}},{{},{{}},{{},{{}}}}\
}},{{},{{}},{{},{{}}},{{},{{}},{{},{{}}}},{{},{{}},{{},{{}}},{{},{{}},{{},{{}}\
}}},{{},{{}},{{},{{}}},{{},{{}},{{},{{}}}},{{},{{}},{{},{{}}},{{},{{}},{{},{{}\
}}}}}},{{},{{}},{{},{{}}},{{},{{}},{{},{{}}}},{{},{{}},{{},{{}}},{{},{{}},{{},\
{{}}}}},{{},{{}},{{},{{}}},{{},{{}},{{},{{}}}},{{},{{}},{{},{{}}},{{},{{}},{{}\
,{{}}}}}},{{},{{}},{{},{{}}},{{},{{}},{{},{{}}}},{{},{{}},{{},{{}}},{{},{{}},{\
{},{{}}}}},{{},{{}},{{},{{}}},{{},{{}},{{},{{}}}},{{},{{}},{{},{{}}},{{},{{}},\
{{},{{}}}}}}}},{{},{{}},{{},{{}}},{{},{{}},{{},{{}}}},{{},{{}},{{},{{}}},{{},{\
{}},{{},{{}}}}},{{},{{}},{{},{{}}},{{},{{}},{{},{{}}}},{{},{{}},{{},{{}}},{{},\
{{}},{{},{{}}}}}},{{},{{}},{{},{{}}},{{},{{}},{{},{{}}}},{{},{{}},{{},{{}}},{{\
},{{}},{{},{{}}}}},{{},{{}},{{},{{}}},{{},{{}},{{},{{}}}},{{},{{}},{{},{{}}},{\
{},{{}},{{},{{}}}}}}},{{},{{}},{{},{{}}},{{},{{}},{{},{{}}}},{{},{{}},{{},{{}}\
},{{},{{}},{{},{{}}}}},{{},{{}},{{},{{}}},{{},{{}},{{},{{}}}},{{},{{}},{{},{{}\
}},{{},{{}},{{},{{}}}}}},{{},{{}},{{},{{}}},{{},{{}},{{},{{}}}},{{},{{}},{{},{\
{}}},{{},{{}},{{},{{}}}}},{{},{{}},{{},{{}}},{{},{{}},{{},{{}}}},{{},{{}},{{},\
{{}}},{{},{{}},{{},{{}}}}}}}}},{{},{{}},{{},{{}}},{{},{{}},{{},{{}}}},{{},{{}}\
,{{},{{}}},{{},{{}},{{},{{}}}}},{{},{{}},{{},{{}}},{{},{{}},{{},{{}}}},{{},{{}\
},{{},{{}}},{{},{{}},{{},{{}}}}}},{{},{{}},{{},{{}}},{{},{{}},{{},{{}}}},{{},{\
{}},{{},{{}}},{{},{{}},{{},{{}}}}},{{},{{}},{{},{{}}},{{},{{}},{{},{{}}}},{{},\
{{}},{{},{{}}},{{},{{}},{{},{{}}}}}}},{{},{{}},{{},{{}}},{{},{{}},{{},{{}}}},{\
{},{{}},{{},{{}}},{{},{{}},{{},{{}}}}},{{},{{}},{{},{{}}},{{},{{}},{{},{{}}}},\
{{},{{}},{{},{{}}},{{},{{}},{{},{{}}}}}},{{},{{}},{{},{{}}},{{},{{}},{{},{{}}}\
},{{},{{}},{{},{{}}},{{},{{}},{{},{{}}}}},{{},{{}},{{},{{}}},{{},{{}},{{},{{}}\
}},{{},{{}},{{},{{}}},{{},{{}},{{},{{}}}}}}}},{{},{{}},{{},{{}}},{{},{{}},{{},\
{{}}}},{{},{{}},{{},{{}}},{{},{{}},{{},{{}}}}},{{},{{}},{{},{{}}},{{},{{}},{{}\
,{{}}}},{{},{{}},{{},{{}}},{{},{{}},{{},{{}}}}}},{{},{{}},{{},{{}}},{{},{{}},{\
{},{{}}}},{{},{{}},{{},{{}}},{{},{{}},{{},{{}}}}},{{},{{}},{{},{{}}},{{},{{}},\
{{},{{}}}},{{},{{}},{{},{{}}},{{},{{}},{{},{{}}}}}}},{{},{{}},{{},{{}}},{{},{{\
}},{{},{{}}}},{{},{{}},{{},{{}}},{{},{{}},{{},{{}}}}},{{},{{}},{{},{{}}},{{},{\
{}},{{},{{}}}},{{},{{}},{{},{{}}},{{},{{}},{{},{{}}}}}},{{},{{}},{{},{{}}},{{}\
,{{}},{{},{{}}}},{{},{{}},{{},{{}}},{{},{{}},{{},{{}}}}},{{},{{}},{{},{{}}},{{\
},{{}},{{},{{}}}},{{},{{}},{{},{{}}},{{},{{}},{{},{{}}}}}}}}}}}
\end{verbatim}
and so on.

\smallskip

\begin{definition}
\end{definition}
\emph{sum of natural numbers:}

$ + \in \mathbb{N}^{\mathbb{N}^{2}} $:
\begin{equation}
    n + 0 \; := n
\end{equation}
\begin{equation}
    n + 1 \; := \; Suc(n)
\end{equation}
\begin{equation}\
    n + ( m+1) \; := \; ( n+m)+1
\end{equation}

\smallskip

\begin{definition}
\end{definition}
\emph{product of natural numbers:}

$ \cdot \in \mathbb{N}^{\mathbb{N}^{2}} $:

\begin{equation}
    n \cdot 0 \; := 0
\end{equation}
\begin{equation}
    n \cdot ( m+1) \; := \; ( n \cdot m) + n
\end{equation}

\newpage

\begin{definition}
\end{definition}
\emph{exponentiation of natural numbers:}

$ \cdot ^{ \cdot} \in \mathbb{N}^{\mathbb{N}^{2}}  $:

\begin{equation}
    n^{0} \; := \; 1
\end{equation}
\begin{equation}
    n^{m+1} \; := \; n^{m} \cdot n
\end{equation}

\smallskip
\begin{definition}
\end{definition}
\emph{ordering of natural numbers:}

$ < , \leq \in \mathcal{P} ( \mathbb{N}^{2} ) $:

\begin{equation}
    m < n \; := \; m \in n
\end{equation}
\begin{equation}
    m \leq n \; := \; m \subset n
\end{equation}

\bigskip

\begin{definition} \label{def:set of all integer numbers}
\end{definition}
\emph{set of all integer numbers:}
\begin{equation}
    \mathbb{Z} \; := \; \frac{ \mathbb{N}^{2} }{\sim_{\mathbb{Z}}}
\end{equation}
where $ \sim_{\mathbb{Z}} $ is the following equivalence relation
over $ \mathbb{N}^{2} $:
\begin{equation}
    < n_{1} , m_{1} > \, \sim_{\mathbb{Z}} \, < n_{2} , m_{2} > \; := \;
    n_{1} + m_{2} \, = \,  n_{2} + m_{1}
\end{equation}

\begin{definition}
\end{definition}
\emph{sum of integer numbers:}

$ + \in \mathbb{Z}^{\mathbb{Z}^{2}}$:
\begin{equation}
    [  < n_{1} , m_{1} > ] +  [  < n_{2} , m_{2} > ] \; := \;  [  < n_{1} +  n_{2} , m_{1} +  m_{2} > ]
\end{equation}

\begin{definition}
\end{definition}
\emph{subtraction of integer numbers:}

$ - \in \mathbb{Z}^{\mathbb{Z}^{2}} $:
\begin{equation}
    [  < n_{1} , m_{1} > ] -  [  < n_{2} , m_{2} > ] \; := \;  [  < n_{1} ,  m_{1} >
    ] + [  < m_{2} ,  n_{2} >
    ]
\end{equation}

\begin{definition}
\end{definition}
\emph{product of integer numbers:}

$ \cdot \in \mathbb{Z}^{\mathbb{Z}^{2}} $:
\begin{equation}
    [  < n_{1} , m_{1} > ] \cdot  [  < n_{2} , m_{2} > ] \; := \;  [  < n_{1} \cdot n_{2} + m_{1} \cdot  m_{2} , n_{1} \cdot m_{2} + m_{1} \cdot n_{2}   > ]
\end{equation}

\begin{definition}
\end{definition}
\emph{ordering of integer numbers:}

$ \leq \in \mathcal{P} ( \mathbb{Z}^{2} )$:
\begin{equation}
    [  < n_{1} , m_{1} > ]  \leq  [  < n_{2} , m_{2} > ] \; := \;
    n_{1}+ m_{2} \leq n_{2} + m_{1}
\end{equation}

\bigskip

\begin{definition} \label{def:set of all rational numbers}
\end{definition}
\emph{set of all rational numbers}
\begin{equation}
    \mathbb{Q} \; := \; \{ [ < a , b > ] : a , b \in \mathbb{Z} \wedge b \neq 0 \}
\end{equation}
where the equivalence classes are taken with respect to the
  the following equivalence relation $ \sim_{\mathbb{Q}} $ over
$\mathbb{Z}^{2} $:
\begin{equation}
    < a_{1} , b_{1} > \, \sim_{\mathbb{Q}} \, < a_{2} , b_{2} > \; := \;
   [ (  a_{1} \cdot b_{2} = a_{2} \cdot b_{1} )  \wedge ( b_{1} \neq
   0 \neq b_{2} ) ] \vee [ b_{1} = b_{2} = 0 ]
\end{equation}

\begin{definition}
\end{definition}
\emph{sum of rational numbers:}

$ + \in \mathbb{Q}^{\mathbb{Q}^{2}}$:
\begin{equation}
    [  < a_{1} , b_{1} > ] + [  < a_{2} , b_{2} > ] \; := \; [ <
    a_{1} \cdot b_{2} + a_{2} \cdot b_{1} , b_{1} \cdot b_{2} > ]
\end{equation}

\begin{definition}
\end{definition}
\emph{product of rational numbers:}

$ \cdot \in \mathbb{Q}^{\mathbb{Q}^{2}}$:
\begin{equation}
    [  < a_{1} , b_{1} > ] \cdot [  < a_{2} , b_{2} > ] \; := \; [ <
    a_{1} \cdot a_{2} , b_{1} \cdot b_{2} >]
\end{equation}

\begin{definition}
\end{definition}
\emph{ordering of rational numbers:}

$ \leq \in \mathcal{P} ( \mathbb{Q}^{2} )$:
\begin{equation}
    [  < a_{1} , b_{1} > ] \leq [  < a_{2} , b_{2} > ] \; := \;
    b_{1} \geq 0 \wedge b_{2} \geq 0 \wedge a_{1} \cdot b_{2} \leq
    a_{2} \cdot b_{1}
\end{equation}

\bigskip

\begin{definition} \label{def:unary real interval}
\end{definition}
\emph{unary real interval:}
\begin{equation}
  [ 0 , 1 ] \; := \; \frac{\{ 0 , 1 \}^{\mathbb{N}}}{ \sim_{\mathbb{R}}}
\end{equation}
where $ \sim_{\mathbb{R}} $ is the following equivalence relation
over $ \{ 0 , 1 \}^{\mathbb{N}} $:
\begin{multline}
    \{ a_{n} \}_{n \in \mathbb{N}}  \, \sim_{\mathbb{R}} \,  \{ b_{n} \}_{n \in
    \mathbb{N}} \; := \; \\
    ( \{ a_{n} \}_{n \in \mathbb{N}}  =  \{ b_{n} \}_{n \in
    \mathbb{N}} ) \vee ( \exists n \in \mathbb{N} : \forall k \in
    \mathbb{N} [ ( k < n \Rightarrow a_{k} = b_{k} ) \wedge (
    a_{n} = 1 \wedge b_{n} =0 ) \wedge ( k > n \Rightarrow a_{k}
    =0 \wedge b_{k} = 1 ) ])
\end{multline}

\begin{definition}
\end{definition}
\emph{ordering on $[0,1]$:}

$ \leq \in \mathcal{P} ( [0,1]^{2} )$:
\begin{multline}
    [ \{ a_{n} \}_{n \in \mathbb{N}} ] \leq  [ \{ b_{n} \}_{n \in \mathbb{N}}
    ] \; := [ \{ a_{n} \}_{n \in \mathbb{N}} ] = [ \{ b_{n} \}_{n \in
    \mathbb{N}} ] \vee [ \exists n \in \mathbb{N} : a_{n} < b_{n}
   \, \wedge \, ( a_{k} = b_{k} \; \forall k \in n  )]
\end{multline}

\begin{definition}
\end{definition}
\emph{set of all real numbers:}
\begin{equation}
    \mathbb{R} \; := \; \mathbb{Z} \times [ 0 , 1)
\end{equation}
where:
\begin{equation}
  [ 0 , 1) \; := [ 0 ,1 ] - \{ 1 \}
\end{equation}

\begin{definition}
\end{definition}
\emph{ordering on $ \mathbb{R} $:}

$ \leq \in \mathcal{P} ( \mathbb{R}^{2} )$:
\begin{equation}
    < k , r > \, \leq \, < l , s > \; := \; k < l \, \vee \, ( k = l \wedge  r \leq
    s )
\end{equation}

\bigskip

Let us now briefly review how the theory of of ordinal and
cardinal numbers is introduced within ZFC.

\begin{definition}
\end{definition}
\emph{partially ordered set:}

a couple $ ( S , \leq ) $ such that:
\begin{enumerate}
    \item S is a set
    \item $ \leq $ is a partial ordering over S
\end{enumerate}

\begin{definition}
\end{definition}
\emph{totally ordered set:}

a couple $ ( S , \leq ) $ such that:
\begin{enumerate}
    \item S is a set
    \item $ \leq $ is a total ordering over S
\end{enumerate}

\begin{definition}
\end{definition}
\emph{well-ordered set:}

a couple $ ( S , \leq ) $ such that:
\begin{enumerate}
    \item S is a set
    \item $ \leq $ is a well-ordering over S
\end{enumerate}

Given two partially ordered sets $ ( S_{1} , \leq_{1} ) $ and $ (
S_{2} , \leq_{2} ) $:
\begin{definition}
\end{definition}
\emph{$ ( S_{1} , \leq_{1} ) $ and $ ( S_{2} , \leq_{2} ) $ have
the same order-type:}
\begin{equation}
   ( S_{1} , \leq_{1} ) \, \sim_{ord} \,  ( S_{2} , \leq_{2} ) \; := \; \exists f \in S_{2}^{S_{1}} \, bijective \,  : \; [( x \leq_{1} y \;
    \Leftrightarrow \; f(x) \leq_{2} f(y) ) \, \forall x,y \in
    S_{1} ]
\end{equation}

Given a set $ \alpha $:
\begin{definition}
\end{definition}
\emph{$ \alpha $ is an ordinal number:}
\begin{enumerate}
    \item
\begin{equation}
    \beta \in \alpha \; \Rightarrow \; \beta \subset \alpha
\end{equation}
    \item
\begin{equation}
    ( \beta = \gamma \, \vee \, \beta \in \gamma \, \vee \, \gamma
    \in \beta ) \; \; \forall \beta , \gamma \in \alpha
\end{equation}
    \item
\begin{equation}
    \emptyset \neq \beta \subset \alpha \; \Rightarrow \; \exists
    \gamma \in \beta  \, : \, \gamma \cap \beta = \emptyset
\end{equation}
\end{enumerate}

Let us furnish, for completeness, also the following
\cite{Knuth-74}, \cite{Conway-01}, \cite{Gonshor-86}:
\begin{definition}
\end{definition}
\emph{$ \alpha $ is a surreal number:}
\begin{equation}
    \exists \beta \text{ ordinal number } \; : \; \alpha \in \{ 0
    , 1\}^{\beta}
\end{equation}

\begin{theorem}
\end{theorem}
\begin{enumerate}
    \item
\begin{equation}
    n \text{ is an ordinal number } \; \; \forall n \in \mathbb{N}
\end{equation}
    \item
\begin{equation}
     \mathbb{N} \text{ is an ordinal number }
\end{equation}
    \item
\begin{equation}
\alpha + 1 \; := \; Suc( \alpha ) \text{ is an ordinal number } \;
\; \forall \alpha \text{ ordinal number}
\end{equation}
   \item
\begin{equation}
    ( \alpha , \leq ) \text{ is a well-ordered set}  \;
\; \forall \alpha \text{ ordinal number }
\end{equation}
where $ \leq \; := \; \subset $.
   \item
\begin{equation}
   [  ( \alpha , \leq )  \sim_{ord} ( \beta , \leq ) \;
   \Rightarrow \; \alpha = \beta ] \; \; \forall \alpha , \beta \text{ ordinal numbers }
\end{equation}
 \item
\begin{equation}
  ( \alpha = \beta \; \vee \; \alpha < \beta \; \vee \; \beta <
  \alpha ) \; \; \forall \alpha , \beta \text{ ordinal numbers }
\end{equation}
where $ < \; := \; \in $.
\end{enumerate}

Given an ordinal number $ \alpha $:
\begin{definition}
\end{definition}
\emph{$ \alpha $ is an ordinal successor:}
\begin{equation}
    \exists \beta \text{ ordinal number } \; : \; \alpha = Suc(
    \beta )
\end{equation}
\begin{definition}
\end{definition}
\emph{$ \alpha $ is a limit ordinal:}
\begin{equation}
    \nexists \beta \text{ ordinal number } \; : \; \alpha = Suc(
    \beta )
\end{equation}

\begin{example}
\end{example}
Every $ n \in \mathbb{N} $ as well as $ \mathbb{N} +1 , \cdots ,
\mathbb{N} + n $ are ordinal successors.

$ \mathbb{N} $, instead,  is a limit ordinal.

\newpage
\begin{theorem}
\end{theorem}
\emph{ordinal numbers as demarcators of order-type:}

\begin{hypothesis}
\end{hypothesis}
\begin{center}
 $ ( W , \leq )$ well-ordered set
\end{center}
\begin{thesis}
\end{thesis}
\begin{equation}
    \exists !  \, \alpha \text{ ordinal number } \; : \; ( W , \leq )
    \sim_{ord} ( \alpha , \subset )
\end{equation}

\smallskip

Given two sets $ S_{1} $ and $ S_{2} $:
\begin{definition} \label{def:have the same cardinality}
\end{definition}
\emph{$ S_{1} $ and $ S_{2} $ have the same cardinality:}
\begin{equation}
   S_{1} \sim_{card} S_{2} \; := \;  \exists f \in S_{2}^{S_{1}} \; bijective
\end{equation}

Given a set S:
\begin{definition}
\end{definition}
\emph{cardinality of S:}
\begin{equation}
    | S | \; := \; \min \{ \alpha : \alpha  \text{ is an ordinal number
    } \, \wedge \, S \sim_{card} \alpha \}
\end{equation}

The name of the definition \ref{def:have the same cardinality} is
justified by the fact that given two sets $ S_{1} $  and  $ S_{2}
$:
\begin{theorem}
\end{theorem}
\begin{equation}
      S_{1} \sim_{card} S_{2} \; \Leftrightarrow \; | S_{1} |  \,
      = \,  | S_{2} |
\end{equation}

Given a set S:
\begin{definition} \label{def:finite set}
\end{definition}
\emph{S is finite:}
\begin{equation}
     | S | \in \mathbb{N}
\end{equation}

\begin{definition} \label{def:infinite set}
\end{definition}
\emph{S is infinite:}
\begin{equation}
  | S | \notin \mathbb{N}
\end{equation}

\begin{definition}
\end{definition}
\emph{S is countable:}
\begin{equation}
  | S | \; = \; \mathbb{N}
\end{equation}

\begin{definition}
\end{definition}
\emph{S is uncountable:}
\begin{equation}
  | S | \; > \; \mathbb{N}
\end{equation}

\begin{definition}
\end{definition}
\emph{S has the cardinality of the continuum:}
\begin{equation}
    |S| \; = \; | \mathbb{R} |
\end{equation}

Let us introduce the following:
\begin{definition} \label{def:ZFC+CH}
\end{definition}
\emph{formal system ZFC+CH:}

the formal system ZFC augmented with the following:
\begin{axiom} \label{ax:CH}
\end{axiom}
\emph{Continuum Hypothesis (CH):}
\begin{equation}
   | \mathbb{R} | \; = \; | \mathcal{P} ( \mathbb{N}) |
\end{equation}

\begin{remark}
\end{remark}
The Continuum Hypothesis (i.e. axiom \ref{ax:CH}) is rather
intuitive since it states that there don't exist "intermediate
degrees" of infinity between the discrete (having the cardinality
of  $ \mathbb{N} $) and the continuum (having the cardinality of $
\mathbb{R}$).

The reason why one has to add it as a new axiom is that Paul Cohen
has proved that it is undecidable (i.e. it can be neither proved
nor disproved) within ZFC.

G\"{o}del has, anyway, proved that ZFC+CH is consistent relative
to ZFC, i.e. that if ZFC is consistent it follows that ZFC+CH is
consistent too.

\smallskip

We will assume that the formal system giving foundation to
Mathematics is ZFC+CH that, in particular, we will assume in this
paper.

\begin{remark}
\end{remark}
It may be appropriate to conclude this section with a peroration
in favor of the ZFC-orthodoxy.

The foundation of Mathematics given by ZFC has passed the test of
nearly a century with excellent results.

Though, according to G\"{o}del's Second Theorem
\cite{Odifreddi-89}, we cannot prove the consistency of ZFC  from
within ZFC itself, no inconsistency in it has been found.

So why to give up to it ?

Yes, it is true that within ZFC one cannot consider proper classes
and extend by them the Comprehension Scheme (such as in the Von
Neumann Bernays-G\"{o}del augmented with the Axiom of Choice
(VNBGZ) formal system \cite{Jech-02}, but is this something really
useful considering that provability in ZFC implies provability in
VNBGC and Schoenfeld's Theorem stating that a sentence involving
only set variables provable in VNBG is provable also in ZF ?

Or  hasn't Peter Aczel's Theory of Hypersets (see for instance the
Appendix B "Axioms and Universes" of \cite{Moschovakis-94})
obtained giving up the Axiom of Foundation by allowing the
membership relation $ \in $ to be not well-founded (well-foundness
condition that can be rephrased as the assumption that decorations
are defined only for trees) in the more radical way (i.e.
assuming, through the Axiom of Antifoundation, that every graph
can be decorated) simply put us again on the edge of Russell's
abyss allowing the case in which $ x \in x $ for exoterical
intrinsically non-recursive mathematical objects such as the
hyperset $ x := Suc(x) $, implemented by the Mathematica
expression:
\begin{verbatim}
$RecursionLimit=Infinity;

 x := { x , {x}  }
\end{verbatim}
obviously non-halting ?

Or has the alternative topos-theoretic foundation of Mathematics
given by the first-order theory (WPT) of well-pointed topoi
presented some concrete advantages balancing the discouragement of
having to handle intuitionistic logic and the fact that WPT hasn't
the full strength of ZFC as it is shown by the fact that it is
equiconsistent only with the weaker formal system RZC of
Restricted Zermelo set theory  augmented with the Axiom of Choice
in which constraints are posed on the adoption of quantifiers in
the Comprehension Axiom (see the $ 10^{th} $ section "Topos
Theoretic and Set Theoretic Foundations" of the $ 6^{th} $ chapter
"Topoi and Logic" of \cite{Mc-Lane-Moerdijk-92}) ?

\smallskip

Experience have taught us to defend ZFC's orthodoxy.

\smallskip

As to extensions of ZFC, while the passage from ZFC to ZFC+CH has
not bad consequences, there exist pernicious extensions of ZFC
with catastrophic consequences, as we will show later.

\newpage
\section{Nonstandard Analysis remaining inside ZFC+CH} \label{sec:Nonstandard Analysis remaining inside ZFC+CH}
In this section we present a brief review of the orthodox
foundations of Nonstandard Analysis given remaining within  the
formal system ZFC+CH of definition \ref{def:ZFC+CH}
\cite{Robinson-96}, \cite{Goldblatt-98}, \cite{Davis-05}.

 Given a set $S \neq \emptyset $:
\begin{definition}
\end{definition}
\emph{filter on S:}
\begin{equation}
    \mathcal{F} \subseteq \mathcal{P} (S) \, : \, ( A \cap B \in
    \mathcal{F} \; \; \forall A , B \in \mathcal{F} ) \: \wedge \: [( A
    \in \mathcal{F}) \, \wedge \,( A \subseteq B \subseteq S \, \Rightarrow
    \, B \in \mathcal{F} )]
\end{equation}
\begin{definition}
\end{definition}
\emph{ultrafilter on S:}

a filter $  \mathcal{F} $ on S such that:
\begin{equation}
   \mathcal{F} \neq \mathcal{P} (S) \; \wedge \; ( A \in
   \mathcal{F} \vee S - A \in \mathcal{F} \; \; \forall A \in  \mathcal{P}
   (S) )
\end{equation}

Given $ B \in \mathcal{P} (S) : B \neq \emptyset $:
\begin{definition}
\end{definition}
\emph{principal filter generated by  B:}
\begin{equation}
    \mathcal{F}^{B} \; := \; \{ A \in \mathcal{P} (S) \, : \, A \supseteq B \}
\end{equation}

A consequence of the Axiom of Choice (i.e axiom \ref{ax:choice})
is the following \cite{Goldblatt-98}:

\begin{proposition} \label{prop:exists non principal ultrafilter over any infinite set}
\end{proposition}
\begin{equation}
    | S | \geq | \mathbb{N} | \; \Rightarrow \; \exists  \, \mathcal{F} \:
    \text {nonprincipal ultrafilter on S}
\end{equation}

Given $ \bar{r} = \{ r_{n} \}_{n \in \mathbb{N}} ,  \bar{s} = \{
s_{n} \}_{n \in \mathbb{N}} \in \mathbb{R}^{\mathbb{N}}$:

\begin{definition}
\end{definition}
\begin{equation}
    \bar{r} \oplus \bar{s} \; := \;  \{ r_{n} + s_{n} \}_{n \in \mathbb{N}}
\end{equation}
\begin{equation}
    \bar{r} \odot \bar{s} \; := \;  \{ r_{n} \cdot s_{n} \}_{n \in \mathbb{N}}
\end{equation}

\newpage
Given $ x \in \mathbb{R} $:
\begin{definition}
\end{definition}
\begin{equation}
    x^{\mathbb{N}} \; := \; \text{ the only element of } \{ x \}^{\mathbb{N}}
\end{equation}

Let us now introduce the following:
\begin{definition}
\end{definition}
\begin{equation}
    NPU( \mathbb{N} ) \; := \; \{ \mathcal{F} \:
    \text {nonprincipal ultrafilter on } \mathbb{N} \}
\end{equation}

By Proposition \ref{prop:exists non principal ultrafilter over any
infinite set} it follows that:
\begin{proposition}
\end{proposition}
\begin{equation}
    NPU( \mathbb{N} )  \; \neq \; \emptyset
\end{equation}

Given $ \mathcal{F} \in  NPU( \mathbb{N} ) $ and $ \bar{r} = \{
r_{n} \}_{n \in \mathbb{N}} ,  \bar{s} = \{ s_{n} \}_{n \in
\mathbb{N}} \in \mathbb{R}^{\mathbb{N}}$:
\begin{definition}
\end{definition}
\emph{$ \bar{r} $ and  $ \bar{s} $ are equal $ \mathcal{F}$-almost
everywhere:}
\begin{equation}
 \bar{r} \sim_{\mathcal{F}} \bar{s} \; := \; \{ n \in \mathbb{N} \, : \, r_{n} = s_{n}
 \} \in \mathcal{F}
\end{equation}

It may be proved that \cite{Goldblatt-98}:
\begin{proposition}
\end{proposition}
\begin{center}
  $ \sim_{\mathcal{F}} $ is an equivalence relation over $
  \mathbb{R}^{\mathbb{N}}$
\end{center}

Let us finally introduce the following:
\begin{definition}
\end{definition}
\begin{equation}
    ^{\star} \mathbb{R}_{\mathcal{F}} \; = \; \frac{\mathbb{R}^{\mathbb{N}}}{\sim_{\mathcal{F}}}
\end{equation}

Given $ \bar{r} = \{ r_{n} \}_{n \in \mathbb{N}} ,  \bar{s} = \{
s_{n} \}_{n \in \mathbb{N}} \in \mathbb{R}^{\mathbb{N}}$:
\begin{definition}
\end{definition}
\begin{enumerate}
    \item
 \begin{equation}
    [ \bar{r} ]_{\mathcal{F}} + [ \bar{s} ]_{\mathcal{F}} \; := \;
    [ \bar{r} \oplus  \bar{s} ]_{\mathcal{F}}
\end{equation}
    \item
\begin{equation}
    [ \bar{r} ]_{\mathcal{F}} \cdot [ \bar{s} ]_{\mathcal{F}} \; := \;
    [ \bar{r} \odot  \bar{s} ]_{\mathcal{F}}
\end{equation}
    \item
\begin{equation}
    [ \bar{r} ]_{\mathcal{F}} \leq [ \bar{s} ]_{\mathcal{F}} \; := \;
    \{ n \in \mathbb{N} \, : \, r_{n} \leq s_{n} \} \in {\mathcal{F}}
\end{equation}
\end{enumerate}

\smallskip

The assumption of the Continuum Hypothesis (i.e. axiom
\ref{ax:CH}) implies that:
\begin{proposition} \label{prop:independence from the nonprincipal ultrafilter}
\end{proposition}
\begin{center}
   $ (  ^{\star} \mathbb{R}_{\mathcal{F}_{1}} , + , \cdot , \leq )
   $ is isomorphic to $ (  ^{\star} \mathbb{R}_{\mathcal{F}_{2}} , + , \cdot , \leq
   )  \; \; \forall \mathcal{F}_{1} ,  \mathcal{F}_{2} \in NPU( \mathbb{N}
  ) $
\end{center}

Proposition \ref{prop:independence from the nonprincipal
ultrafilter} allows to give the following:

\begin{definition} \label{def:hyperreal number system of Nonstandard Analysis}
\end{definition}
\emph{hyperreal number system of Nonstandard Analysis:}
\begin{equation}
    (  ^{\star} \mathbb{R} , + , \cdot , \leq ) \; := \; ( ^{\star} \mathbb{R}_{\mathcal{F}} , + , \cdot , \leq
    ) \; \;  \mathcal{F} \in NPU( \mathbb{N})
\end{equation}

\smallskip

Given $ x \in \, ^{\star} \mathbb{R} $:

\begin{definition}
\end{definition}
\begin{equation}
    | x | \; := \; \left\{%
\begin{array}{ll}
    x, & \hbox{ if $ x \geq 0 $;} \\
    -x, & \hbox{if $ x < 0$.} \\
\end{array}%
\right.
\end{equation}

\begin{definition}
\end{definition}
\emph{x is infinitesimal:}
\begin{equation}
    | x | < \epsilon \; \; \forall \epsilon \in ( 0 , + \infty
    )
\end{equation}
\begin{definition}
\end{definition}
\emph{x is limited:}
\begin{equation}
    \exists r \in \mathbb{R} \; : \;  | x | < r
\end{equation}
\begin{definition}
\end{definition}
\emph{x is unlimited:}
\begin{equation}
    | x | > r \; \; \forall r \in ( 0 , + \infty
    )
\end{equation}

Given $ x_{1}, x_{2} \in \, ^{\star} \mathbb{R} $:
\begin{definition}
\end{definition}
\emph{$ x_{1} $ is infinitely closed to $ x_{2} $:}
\begin{equation}
    x_{1} \simeq x_{2} \; := \; x_{1}-x_{2} \text{ is infinitesimal }
\end{equation}

It can be easily verified that:
\begin{proposition}
\end{proposition}
\begin{center}
  $ \simeq $ is an equivalence relation over $  ^{\star}
  \mathbb{R} $.
\end{center}

Given $ x \in \,  ^{\star} \mathbb{R} $:
\begin{definition}
\end{definition}
\emph{halo of x:}
\begin{equation}
    hal( x ) \; := \; \{ y \in  \, ^{\star}
  \mathbb{R} : x \simeq y    \}
\end{equation}

Clearly the set of infinitesimal hyperreals is nothing but $
hal(0) $.

\newpage

\begin{proposition} \label{prop:exhibition of some infinitesimal and unlimited hyperreals}
\end{proposition}

\begin{enumerate}
    \item
\begin{equation}
    [ \bar{r}]_{\mathcal{F}} \in hal(0) \; \; \forall \bar{r} = \{ r_{n} \}_{n \in
    \mathbb{N}} \in \mathbb{R}^{\mathbb{N}} \, : \, \lim_{n \rightarrow +
    \infty} r_{n} = 0
\end{equation}
    \item
\begin{equation}
      [ \bar{r}]_{\mathcal{F}} \text{ is unlimited } \; \; \forall \bar{r} = \{ r_{n} \}_{n \in
    \mathbb{N}} \in \mathbb{R}^{\mathbb{N}} \, : \, \lim_{n \rightarrow +
    \infty} r_{n} = + \infty
\end{equation}
\end{enumerate}

\begin{remark} \label{rem:hyperreals and constructivism}
\end{remark}
Nonstandard Analysis is often criticized for its claimed
nonconstructive nature.

Alain Connes, for instance, claims in \cite{Connes-98} that no
element of $ ^{\star} \mathbb{R} - \mathbb{R} $ "can be
exhibited".

So a short confutation of Connes' claim based on proposition
\ref{prop:exhibition of some infinitesimal and unlimited
hyperreals} could simply  be the exhibition of the infinitesimal
hyperreal $ [ \{ \frac{1}{n} \}_{n \in \mathbb{N}} ]_{\mathcal{F}}
$ or of the unlimited hyperreal $ [ \{ n \}_{n \in \mathbb{N}}
]_{\mathcal{F}} $.

A more detailed analysis of Connes' statement requires, anyway, a
precise definition of the locution "exhibiting a mathematical
object" and consequentially naturally  leads us to the issue of
furnishing a precise definition of the term "constructive".

As we have implicitly done in the remark \ref{rem:nonconstructive
nature of the Axiom of Foundation} and in the remark
\ref{rem:nonconstructive nature of the Axiom of Choice} we will
say that the proposition stating the existence of a mathematical
object x is constructive whether is contains also the explicit
definition of an algorithm, i.e. (assuming Church Thesis) a
partial recursive function \cite{Cutland-80}, \cite{Odifreddi-89},
computing x.

With this regard the definition \ref{def:hyperreal number system
of Nonstandard Analysis} of the hyperreals as $
\sim_{\mathcal{F}}$-equivalence classes of real sequences is as
much constructive as the definition \ref{def:unary real interval}
of the reals belonging to the interval $[ 0,1]$ as $
\sim_{\mathbb{R}} $-equivalence classes of binary sequences.

The fact that the former is based on the nonconstructive Axiom of
Choice is balanced by the fact the latter is (implicitly) based on
the nonconstructive Axiom of Foundation.

\bigskip

Given $ x \in \, ^{\star} \mathbb{R} $ limited:
\begin{proposition} \label{prop:existence of the standard part of a limited hyperreal}
\end{proposition}
\begin{equation}
    \exists \, ! \, st(x) \in \mathbb{R} \; : \; st(x) \simeq x
\end{equation}

\begin{proposition} \label{prop:existence of the standard part as Dedekind completeness}
\end{proposition}
\begin{center}
 Proposition \ref{prop:existence of the standard part of a limited
 hyperreal} is equivalent to the Dedekind completeness of $
 \mathbb{R}$
\end{center}

st(x) is called the standard part of the limited hyperreal x.

\smallskip

Given a set $ A \in \mathcal{P} ( \mathbb{R} ) $:
\begin{definition}
\end{definition}
\emph{enlargement of A:}
\begin{equation}
    ^{\star} A \: := \; \{ [ \bar{r} ]_{\mathcal{F}} \in \,
    ^{\star} \mathbb{R} \, : \, \{ n \in \mathbb{N} : r_{n} \in A
    \} \in \mathcal{F} \}
\end{equation}

\begin{example}
\end{example}
$ ^{\star} \mathbb{N} $ is usually called the set of all
hypernatural numbers, $ ^{\star} \mathbb{Z} $ is usually called
the set of all hyperinteger numbers and $ ^{\star} \mathbb{Q} $ is
usually called the set of all hyperrational numbers.

\smallskip
Given a set S:
\begin{definition} \label{def:hyperfinite set}
\end{definition}
\emph{S is hyperfinite:}
\begin{equation}
    \exists n \in \, ^{\star} \mathbb{N} \; : \; S \, = \, \{ k
    \in \, ^{\star} \mathbb{N} : k \leq n \}
\end{equation}

\begin{remark}
\end{remark}
Let us remark that clearly, according to definition
\ref{def:infinite set}, an hyperfinite set is infinite.

\smallskip

Given $ a , b \in \mathbb{R} : a < b $ and an hypernatural $ n \in
\, ^{\star} \mathbb{N} $:
\begin{definition}
\end{definition}
\emph{n-sliced interval between a and b:}
\begin{equation}
    [ a , b ]_{n} \; := \; \{  a + k \cdot \frac{b-a}{n} \; k \in \{ j \in \, ^{\star} \mathbb{N} : j \leq n  \}     \}
\end{equation}

\begin{definition}
\end{definition}
\emph{$[ a , b ]_{n}$ is an hyperfinite interval:}
\begin{equation}
    n  \; \in \; ^{\star} \mathbb{N} - \mathbb{N}
\end{equation}

\begin{remark}
\end{remark}
Let us remark that clearly, according to definition
\ref{def:infinite set}, an hyperfinite interval is infinite.

\bigskip

In order to introduce some more advanced technique of Nonstandard
Analysis it is useful to introduce some new set-theoretic notion.

Given a set S and $ n \in \mathbb{N} $:
\begin{definition}
\end{definition}
\emph{ $ n^{th} $ cumulative power set of S:}
\begin{equation}
    \mathbb{U}_{0} (S) \; := \; S
\end{equation}
\begin{equation}
   \mathbb{U}_{n} (S) \; := \; \mathbb{U}_{n-1} (S) \, \cup \,
   \mathcal{P} (  \mathbb{U}_{n-1} (S) )
\end{equation}

The cumulative power sets of finite sets may be computed through
the following Mathematica \cite{Wolfram-96} code:
\begin{verbatim}
$RecursionLimit=Infinity;

<<DiscreteMath`Combinatorica`

powerset[x_]:=LexicographicSubsets[x]

cumulativepowerset[x_,n_]:=
  If[n==0,x,Union[cumulativepowerset[x,n-1],powerset[cumulativepowerset[x,n-1]]]]

\end{verbatim}

\begin{example}
\end{example}
Let us compute the first cumulative power sets of the empty set.
We obtain that:
\begin{verbatim}
cumulativepowerset[{},0 ] = {}

cumulativepowerset[{},1 ] = {{}}

cumulativepowerset[{},2 ] = {{},{{}}}

cumulativepowerset[{},3 ] = {{},{{}},{{{}}},{{},{{}}}}

cumulativepowerset[{},4 ] =
{{},{{}},{{{}}},{{{{}}}},{{{},{{}}}},{{},{{}}},{{},{{{}}}},{{},{{},{{}}}},{{{}\
},{{{}}}},{{{}},{{},{{}}}},{{{{}}},{{},{{}}}},{{},{{}},{{{}}}},{{},{{}},{{},{{\
}}}},{{},{{{}}},{{},{{}}}},{{{}},{{{}}},{{},{{}}}},{{},{{}},{{{}}},{{},{{}}}}}

\end{verbatim}
and so on.

\begin{definition}
\end{definition}
\emph{superstructure over S:}
\begin{equation}
   \mathbb{U} (S) \; := \; \cup_{n \in \mathbb{N}} \mathbb{U}_{n} (S)
\end{equation}

Substantially every mathematical object needed to study S is an
element of $  \mathbb{U} (S) $.

For instance:
\begin{proposition} \label{prop:the rank of useful ingredients}
\end{proposition}
\begin{enumerate}
    \item the set of all topologies on S is an element of $
    \mathbb{U}_{3}(S)$
    \item the set of all measures on S is an element of $
    \mathbb{U}_{5}(S)$
    \item the set of all metrics on S is an element of $
    \mathbb{U}_{6}(S)$
\end{enumerate}

Given $ x \in \mathbb{U} (S) $:
\begin{definition}
\end{definition}
\emph{rank of x:}
\begin{equation}
    rank(x) \; := \; \min \{ n \in \mathbb{N} \, : \, x \in
    \mathbb{U}_{n} (S) \}
\end{equation}

The basic mathematical object of Nonstandard Analysis is then a
suitably defined extension map $ \star :  \mathbb{U} ( \mathbb{R}
) \mapsto \mathbb{U} ( ^{\star} \mathbb{R} ) $.

Given $ x \in \mathbb{U} ( ^{\star} \mathbb{R} ) $:
\begin{definition}
\end{definition}
\emph{x is internal:}
\begin{equation}
    x \in  \, ^{\star} \mathbb{U} ( \mathbb{R} )
\end{equation}
\begin{definition}
\end{definition}
\emph{x is external:}
\begin{equation}
    x \notin \,  ^{\star} \mathbb{U} ( \mathbb{R} )
\end{equation}

The corner-stone of Nonstandard Analysis is the following:
\begin{proposition} \label{prop:transfer principle}
\end{proposition}
\emph{Transfer Principle:}
\begin{center}
 ($ \phi $ holds in $ \mathbb{U} ( \mathbb{R} ) \; \Leftrightarrow
 \; \star \phi $ holds in $  ^{\star} \mathbb{U} ( \mathbb{R} ) )
 \; \; \forall \phi $ bounded-quantifier statement
\end{center}

A consequence of Proposition \ref{prop:transfer principle} is the
following:
\begin{proposition} \label{prop:conservative property of Nonstandard Analysis}
\end{proposition}
\emph{Conservative Property of Nonstandard Analysis:}
\begin{center}
 Every theorem about $ \mathbb{U} ( \mathbb{R} ) $ that can be
 proved resorting to Nonstandard Analysis (i.e. by using elements of $  \mathbb{U} ( ^{\star}  \mathbb{R}
 )) $  can be also proved without resorting to Nonstandard
 Analysis.
\end{center}

Proposition \ref{prop:conservative property of Nonstandard
Analysis} could lead to think that Nonstandard Analysis is
useless.

It has to be stressed, with this regard, that though every theorem
about $ \mathbb{U} ( \mathbb{R} ) $ that can be
 proved resorting to Nonstandard Analysis can also be proved
 without resorting to it, the complexity of a proof resorting to
 Nonstandard Analysis may be lower than the complexity of a proof
 non resorting to it.

\bigskip

Let us now introduce some notion of Nonstandard Topology.

Given $ M \in \mathbb{U} ( \mathbb{R} ) $ let us recall first of
all that:

\begin{definition}
\end{definition}
\emph{topology over M:}

$\mathcal{T}  \subseteq \mathcal{P} ( M ) $ :
\begin{itemize}
    \item
\begin{equation}
    \emptyset , S \in \mathcal{T}
\end{equation}
    \item
\begin{equation}
    O_{1} , O_{2} \in \mathcal{T} \; \Rightarrow \;  O_{1} \cap O_{2} \in \mathcal{T}
\end{equation}
    \item
\begin{equation}
    O_{i} \in  \mathcal{T}  \,  \forall i \in I \; \Rightarrow \; \cup_{i \in I}
    O_{i} \in \mathcal{T}
\end{equation}
\end{itemize}

We will denote the set of all the topologies over M by TOP(M).

By proposition \ref{prop:the rank of useful ingredients} it
follows that:
\begin{proposition}
\end{proposition}
\begin{equation}
    TOP (M) \in \mathbb{U}_{rank(M)+3}( \mathbb{R} )
\end{equation}

Given $ \mathcal{T} \in TOP(M) $ and $ q_{1}, q_{2} \in M $:
\begin{definition}
\end{definition}
\begin{equation}
    q_{1} \curlyvee q_{2} \; := \; O_{1} \cap  O_{2} \neq \emptyset \; \;  \forall O_{1},O_{2} \in \mathcal{T} \; : \;
  q_{1} \in O_{1} \: \wedge \: q_{2} \in O_{2}
\end{equation}

\begin{definition}
\end{definition}
\emph{$\mathcal{T}$ is Hausdorff:}
\begin{equation}
  \neg ( q_{1} \curlyvee q_{2} ) \; \; \forall q_{1}, q_{2} \in M
\end{equation}

\smallskip

Given $ q_{1} \in M $:

\begin{definition}
\end{definition}
\emph{halo of $ q_{1}$:}
\begin{equation}
    hal(q_{1}) \; := \; \cap_{q_{1} \in O \in \mathcal{T} } \, ^{\star} O
\end{equation}

Then:
\begin{proposition}
\end{proposition}
\begin{equation}
  \mathcal{T}  \text{ is Hausdorff } \; \Leftrightarrow \; hal(
  q_{1}) \cap  hal(q_{2}) \, = \, \emptyset \; \; \forall q_{1} ,
  q_{2} \in M \, : \, q_{1} \neq q_{2} , \forall \mathcal{T} \in
  TOP(M)
\end{equation}

Given $ q_{2} \in \, ^{\star} M $:

\begin{definition}
\end{definition}
\emph{$ q_{2} $ is infinitely closed to $ q_{1}$:}
\begin{equation}
    q_{2} \simeq q_{1} \; := \; q_{2} \in hal( q_{1} )
\end{equation}

\begin{definition}
\end{definition}
\emph{nearstandard points of M:}
\begin{equation}
    ns( M ) \; := \; \{ q_{1} \in \, ^{\star} M \, : \, (  \exists q_{2}
    \in M : q_{1} \simeq q_{2}) \}
\end{equation}

Then:

\begin{proposition} \label{prop:existence of the standard part for Hausdorff topologies}
\end{proposition}

\begin{hypothesis}
\end{hypothesis}
\begin{equation}
    \mathcal{T} \in TOP(M)
\end{equation}
\begin{thesis}
\end{thesis}
\begin{equation}
  \mathcal{T} \text{ is Hausdorff } \; \Rightarrow \; \forall q \in   ns( M ) \,   \exists \, ! \, st(q) \in M \; : \; st(q) \simeq
  q
\end{equation}

\begin{remark}
\end{remark}
Let us consider the particular case in which $ M := \mathbb{R} $
while $ \mathcal{T}_{natural} $ is the natural topology over $
\mathbb{R} $, i.e. the topology induced by the metric
$d_{euclidean}$  induced by the euclidean riemannian metric $
\delta $ over $ \mathbb{R} $.

Then:
\begin{equation}
    ns( \mathbb{R} ) \; = \; \{ x \in \, ^{\star} \mathbb{R} \, :
    \, \text{ x is limited} \}
\end{equation}

Hence, in this case, proposition \ref{prop:existence of the
standard part for Hausdorff topologies} reduces to
 proposition \ref{prop:existence of the standard part of a limited
 hyperreal} that, by proposition \ref{prop:existence of the standard part as Dedekind
 completeness}, we know to be equivalent to the Dedekind
 completeness of $ \mathbb{R} $.

 Consequentially one obtains a different topological viewpoint on
 the Dedekind completeness of $ \mathbb{R} $ that deeply links it
 to the Hausdorffness of $ \mathcal{T}_{natural} $.

\newpage
\section{Nonstandard Analysis going outside ZFC: Internal Set
Theory and its bugs} \label{sec:Nonstandard Analysis going outside
ZFC: Internal Set Theory}

Edward Nelson \cite{Nelson-77} has introduced an alternative
formulation of a part of Nonstandard Analysis  in which the same
formal system ZFC axiomatizing Set Theory is extended to a new
formal system, called Internal Set Theory (shortened as IST),
obtained from ZFC by:
\begin{enumerate}
    \item adding a new undefined unary predicate "standard"
    \item defining an \emph{internal formula} of Internal Set Theory as a formula of ZFC not containing the
    new predicate "standard"
    \item defining an \emph{external formula} of Internal Set
    Theory as a not internal formula
    \item adding to the axioms of ZFC three new
    axioms (the Axiom of Idealization, the Axiom of
    Standardization and the Axiom of Transfer)
\end{enumerate}

\begin{definition}
\end{definition}
\emph{Internal Set Theory (IST):}

the formal system obtained augmenting ZFC with the following:

\begin{axiom}
\end{axiom}
\emph{Axiom of Transfer:}

\begin{hypothesis}
\end{hypothesis}
\begin{center}
  $A( x, t_{1} , \cdots , t_{k} ) $ internal formula with free variables $ x, t_{1} , \cdots , t_{k}$ and no other free variables
\end{center}
\begin{thesis}
\end{thesis}
\begin{equation}
   \forall^{st} t_{1} \cdots \forall^{st} t_{k} ( \forall^{st} x
    \: A( x, t_{1} , \cdots , t_{k} ) \; \Rightarrow \; \forall x
    \,  A( x, t_{1} , \cdots , t_{k} ) )
\end{equation}

\begin{axiom}
\end{axiom}
\emph{Axiom of Idealization:}

\begin{hypothesis}
\end{hypothesis}
\begin{center}
  $B( x, y )$ internal formula with free variables $ x, y $ and possibly other free variables
\end{center}
\begin{thesis}
\end{thesis}
\begin{equation}
   \forall^{st \, fin} z \, \exists x,y \in z : B(x,y) \;
   \Leftrightarrow \; \exists x \, : \, \forall^{st} y \, B(x,y)
\end{equation}
\newpage
\begin{axiom}
\end{axiom}
\emph{Axiom of Standardization:}

\begin{hypothesis}
\end{hypothesis}
\begin{center}
  C(z)  formula, internal or external, with free variable z and possibly other free variables
\end{center}
\begin{thesis}
\end{thesis}
\begin{equation}
   \forall^{st} x \, \exists^{st} y : \forall^{st} z ( z \in y
   \Leftrightarrow z \in x \wedge C(z))
\end{equation}
where we have adopted the following abbreviations:
\begin{equation}
    \forall^{st} x \; := \; \forall x \, : x \; standard
\end{equation}
\begin{equation}
    \exists^{st} x \; := \; \exists x \, : x \; standard
\end{equation}
\begin{equation}
    \forall^{fin} x \; := \; \forall x \, : x \; finite
\end{equation}
\begin{equation}
    \exists^{fin} x \; := \; \exists x \, : x \; finite
\end{equation}

\smallskip

\begin{remark}
\end{remark}
It is important to stress that the fact that the underlying formal
system is different results in that an internal formula $
\phi_{IST} $ of IST has a meaning that is different from the
meaning that the same formula $ \phi_{ZFC} $ has within ZFC.

This implies that the same formal definition of a mathematical
object x results in different mathematical objects according to
whether it is considered within ZFC or within IST;

let us denote  by $ x_{ZFC} $ and by $ x_{IST} $ a mathematical
object x considered within, respectively, the formal system ZFC
and IST.

\begin{remark}
\end{remark}
Since IST is an extension of ZFC it follows that if $ \phi_{ZFC} $
is a theorem of ZFC then $  \phi_{IST} $ is an internal theorem of
IST.

It important, anyway, to stress that $ \phi_{IST} \neq \phi_{ZFC}
$.

\smallskip

For instance the fact that theorem \ref{eq:existence  and unicity
of ZFC's naturals} holds within ZFC implies that a corresponding
theorem holds within IST:

\begin{theorem} \label{eq:existence  and unicity of IFC's naturals}
\end{theorem}
\emph{Existence and unicity of  IST's naturals }
\begin{center}
 There exists exactly one set $  \mathbb{N}_{IST} $ such that:
\begin{enumerate}
    \item
\begin{equation} \label{eq:initial condition in IST}
    \emptyset \in \mathbb{N}_{IST}
\end{equation}
    \item
\begin{equation}  \label{eq:induction condition in IST}
    Suc(x) \in \mathbb{N}_{IST} \; \; \forall x \in \mathbb{N}_{IST}
\end{equation}
where, as in the Axiom of Infinity (i.e axiom \ref{ax:infinity}),
$ Suc(x) := x \cup \{ x \} $.
    \item if K is any set that satisfies eq. \ref{eq:initial
    condition in IST} and eq. \ref{eq:induction condition in IST}  then $
    \mathbb{N}_{IST} \subset K$.
\end{enumerate}
\end{center}
\begin{proof}
By the Axiom of Infinity (i.e axiom \ref{ax:infinity}) there
exists at least one set X satisfying eq. \ref{eq:initial condition
in IST} and eq. \ref{eq:induction condition in IST}.

Let:
\begin{equation}
    \mathcal{F} \; := \; \{ Y \in \mathcal{P} (X) : \emptyset \in Y \, \wedge \, ( Suc(x) \in Y \; \forall x \in Y  \}
\end{equation}
\begin{equation}
    \mathbb{N}_{IST} \; := \; \bigcap_{Y \in  \mathcal{F} } Y
\end{equation}
It is easy to see that the intersection of any nonempty family of
sets satisfying  eq. \ref{eq:initial condition in IST} and eq.
\ref{eq:induction condition in IST} still satisfies  eq.
\ref{eq:initial condition in IST} and eq. \ref{eq:induction
condition in IST}.

Let K be a set that satisfies  eq. \ref{eq:initial condition in
IST} and eq. \ref{eq:induction condition in IST}; then:
\begin{equation}
    X \cap K \in \mathcal{F}
\end{equation}
and:
\begin{equation}
     \mathbb{N}_{IST} \; := \; \bigcap_{Y \in  \mathcal{F} } Y \;
     \subset X \cap K \; \subset \; K
\end{equation}
\end{proof}

It is important anyway to stress that theorem \ref{eq:existence
and unicity of ZFC's naturals} and theorem \ref{eq:existence  and
unicity of IFC's naturals} have different meaning:

the former gives an implicit definition of the set $
\mathbb{N}_{ZFC} $ while the latter gives an implicit definition
of the set $ \mathbb{N}_{IST} $ where:
\begin{equation}
  \mathbb{N}_{ZFC} \neq  \mathbb{N}_{IST}
\end{equation}

Consequentially the chain of mathematical definitions given in the
 section \ref{sec:The orthodox ZFC+CH set-theoretic
foundation of Mathematics} gives rise to sets $ \mathbb{Z}_{IST} $
, $\mathbb{Q}_{IST} $ , $ \mathbb{R}_{IST} $ such that:
\begin{equation}
  \mathbb{Z}_{ZFC} \neq  \mathbb{Z}_{IST}
\end{equation}
\begin{equation}
  \mathbb{Q}_{ZFC} \neq  \mathbb{Q}_{IST}
\end{equation}
\begin{equation}
  \mathbb{R}_{ZFC} \neq  \mathbb{R}_{IST}
\end{equation}

Actually it may be proved that:
\begin{theorem} \label{th:relations between the basic theorems within ZFC and IST}
\end{theorem}
\begin{enumerate}
    \item
\begin{equation}
   \mathbb{N}_{IST} \; = \; ( ^{\star} \mathbb{N} )_{ZFC}
\end{equation}
    \item
\begin{equation}
   \mathbb{Z}_{IST} \; = \; ( ^{\star} \mathbb{Z} )_{ZFC}
\end{equation}
    \item
\begin{equation}
   \mathbb{Q}_{IST} \; = \; ( ^{\star} \mathbb{Q} )_{ZFC}
\end{equation}
    \item
\begin{equation}
   \mathbb{R}_{IST} \; = \; ( ^{\star} \mathbb{R} )_{ZFC}
\end{equation}
\end{enumerate}

The great conceptual bug of Internal Set Theory consists in that:
\newpage
\begin{theorem} \label{th:the set of all ZFC's naturals cannot be defined within IST}
\end{theorem}
\begin{center}
 $ \mathbb{N}_{ZFC} $ cannot be defined within IST
\end{center}
\begin{proof}
 Let us assume ad absurdum that $ ( \mathbb{N} )_{ZFC} $ can be
 defined within IST.

 Then it  satisfies the conditions eq. \ref{eq:initial condition in
IST} and eq. \ref{eq:induction condition in IST} and hence, by
theorem \ref{eq:existence  and unicity of IFC's naturals}, one has
that:
\begin{equation}
  ( \mathbb{N} )_{IST} \; \subset \;  ( \mathbb{N} )_{ZFC}
\end{equation}
that is in contradiction with theorem \ref{th:relations between
the basic theorems within ZFC and IST}
\end{proof}

\begin{remark}
\end{remark}
It is important to remark that within IST given a set S and a
unary predicate p one can use the Comprehension Axiom (i.e. axiom
\ref{ax:comprehension}) to define the set $ S_{p} := \{ x \in S :
p(x) \} $ if and only if p is internal.

Since the predicate:
\begin{equation}
    p_{standard}(x) := \text{ x is standard }
\end{equation}
is not internal it follows that it cannot be used to define the
set $ S_{p_{standard}} $.

This applies in particular for $ S \in \{ \mathbb{N} , \mathbb{Z}
, \mathbb{Q} , \mathbb{R} \} $.

\smallskip

\begin{remark}
\end{remark}
Theorem \ref{th:the set of all ZFC's naturals cannot be defined
within IST} implies that every branch of Mathematics, such as
Arithmetics  or Classical Recursion Theory \cite{Cutland-80},
\cite{Odifreddi-89}, that cannot be formulated without introducing
the set $ ( \mathbb{N} )_{ZFC} $ cannot be formulated within
Internal Set Theory (see  the third chapter "Theories of internal
sets" of \cite{Kanovei-Reeken-04} and in particular the section
3.6c "Three "myths" of IST").

\newpage
\section{Loeb Probability Spaces} \label{sec:Loeb Probability Spaces}

In a remarkable paper \cite{Loeb-75} Peter Loeb introduced a very
rich class of standard measure spaces constructed using
Nonstandard Analysis (in its orthodox formulation given within
ZFC+CH presented in section \ref{sec:Nonstandard Analysis
remaining inside ZFC+CH}). In the later decade Loeb Measures has
demonstrated to be a very powerful tool in the framework of
Classical Probability Theory \cite{Goldblatt-98},
\cite{Cutland-00}.

Let $ \Omega \in \, ^{\star} \mathbb{U} ( \mathbb{R} ) $ be an
internal set, $ \mathcal{A} \subseteq \mathcal{P} ( \Omega ) $ be
an algebra, and $ \mu : \mathcal{A} \rightarrow \,  ^{\star }[ 0 ,
\infty) $ be an internal finitely additive measure on $
\mathcal{A} $ normalized to one, i.e:
\begin{equation}
    \mu ( A_{1} \cup A_{2} ) \; = \; \mu ( A_{1} ) + \mu ( A_{2} )
    \; \; \forall A_{1}, A_{2} \in  \mathcal{A} \, : \, A_{1} \cap
    A_{2} = \emptyset
\end{equation}
\begin{equation}
    \mu ( \Omega ) \; = \; 1
\end{equation}

Given $ B \in \mathcal{P} ( \Omega ) $ (not necessarily internal):
\begin{definition}
\end{definition}
\emph{B is a Loeb $ \mu$-null set:}
\begin{equation}
    \forall \epsilon \in ( 0 , + \infty ) \, , \, \exists A \in \mathcal{A}
    \; : \; B \subseteq A \, \wedge \, \mu (A) < \epsilon
\end{equation}

\begin{definition}
\end{definition}
\emph{Loeb $ \sigma$-algebra w.r.t. $ \mathcal{A} $ and $ \mu $:}
\begin{equation}
    L ( \mathcal{A}, \mu ) \; := \; \{  B \in \mathcal{P} ( \Omega
    ) : (\exists A \in \mathcal{A} : A \triangle B \text{ is a Loeb $ \mu$-null
    set}) \}
\end{equation}
where:
\begin{equation}
  A \triangle B \; := ( A - B ) \cup ( B - A)
\end{equation}
is the symmetric difference of A and B.

\begin{definition}
\end{definition}
\emph{Loeb probability space w.r.t. $ ( \Omega , \mathcal{A} ,
\mu)$: }
\begin{center}
 the classical probability space $  ( \Omega , L (\mathcal{A} , \mu) ,
\mu_{L})$, where $ \mu_{L} : L(A) \mapsto [ 0 , 1] $, said a Loeb
probability measure, is defined as:
\end{center}
\begin{equation}
    \mu_{L} ( A) \; := \; st( \mu (A) ) \; \; A \in  L(A)
\end{equation}

\begin{remark}
\end{remark}
Let us remark that a Loeb probability space is a classical
probability space in the sense of the Kolmogorov axiomatization
\cite{Kolmogorov-56}, \cite{Billingsley-95}.

\smallskip

A particularly important example of a Loeb probability measure is
the Loeb counting measure we are going to introduce.

Given $ n \in \, ^{\star} \mathbb{N} - \mathbb{N} $ let us
consider the hyperfinite set:
\begin{equation}
    \Omega \; := \; \{ j \in \, ^{\star} \mathbb{N} \, : \, 1 \leq j \leq n \}
\end{equation}
and the map $ \nu : \, ^{\star} \mathcal{P} ( \Omega ) \mapsto \,
^{\star} [0,1] $:
\begin{equation}
    \nu (A) \; := \; \frac{^{\star} | A |}{n}
\end{equation}
where $ ^{\star} | \cdot |  $ is the extension to $ ^{\star}
\mathbb{U} ( \mathbb{R} ) $ of the function $ | \cdot | $ that
gives the cardinality of finite sets.

\begin{definition}
\end{definition}
\emph{$ n^{th} $ counting Loeb probability space}
\begin{center}
    the Loeb probability space $ ( \Omega , L[  \, ^{\star} \mathcal{P} ( \Omega
    ) , \nu ] , \nu_{L} ) $.
\end{center}

\end{document}